%% file: arXiv.tex
\documentclass[11pt]{article}
\usepackage[margin=1in]{geometry}  %

\usepackage{graphicx}
\usepackage{upgreek}
\usepackage{multicol,multirow}
\usepackage{amsmath,amssymb,amsfonts}
\usepackage{mathrsfs}
\usepackage{amsthm}
\usepackage[figuresright]{rotating}
\usepackage{appendix}
\usepackage[authoryear]{natbib}
\usepackage{ifpdf}
\usepackage[T1]{fontenc}
\usepackage{newtxtext}
\usepackage{newtxmath}
\usepackage{textcomp}
\usepackage{xcolor}
\usepackage{float}
\usepackage{siunitx}
\usepackage[colorlinks,allcolors=blue]{hyperref}
\usepackage[most]{tcolorbox}

\newtcolorbox{boxtext}{
  colback=gray!10,    %
  colframe=black,     %
  boxrule=0.5pt,      %
  arc=2mm,            %
  left=6pt, right=6pt, top=6pt, bottom=6pt,
}

\newenvironment{Backmatter}{\par\small}{\par}

\definecolor{jourcolor}{cmyk}{1,0.57,0.01,0.38}
\hypersetup{
    colorlinks,
    citecolor=jourcolor,
    filecolor=jourcolor,
    linkcolor=jourcolor,
    urlcolor=jourcolor
}

\theoremstyle{definition}

\title{Simulating The Urban Canopy's Impact on Wind-Driven Natural Ventilation}

\author{
Nicholas Bachand\textsuperscript{1}, 
Hesam Salehipour\textsuperscript{2}, 
Catherine Gorlé\textsuperscript{1} \\
\\
\textsuperscript{1}Civil and Environmental Engineering Department, Stanford University, Stanford, USA \\
\textsuperscript{2}Autodesk Research, Toronto, Canada
}

\date{}

\begin{document}

\maketitle


\vspace{1em}

\input{body}

\bibliography{references}

\end{document}

%% file: body.tex
\abstract{The urban canopy affects wind in complex ways, making it challenging to predict wind-driven natural ventilation and cooling in buildings. Using large eddy simulations of coupled outdoor and indoor airflow, we study how the surrounding urban canopy and wind angle influence ventilation rates through four ventilation configurations: cross, corner, dual-room, and single-sided. Flow visualizations demonstrate how both large-scale flow patterns and local interference effects can influence ventilation rates by 50-85\%. In general, lower density canopies give higher ventilation rates and wind angles that align with a direct path between two openings also lead to higher ventilation rates. However, interference effects from surrounding buildings can significantly change the local wind speed and direction, thus also changing ventilation rates. The magnitude of these interference effects depends on both the wind angle and surrounding building geometries. The effect of wind angle is less pronounced in a higher density canopy, where the urban canopy geometry more strongly guides the flow. The results demonstrate that the canopy's effect on ventilation rates is much more complex than those suggested by existing natural ventilation parameterizations.}

\maketitle

\begin{boxtext}

\textbf{\mathversion{bold}Impact Statement}

The passive cooling of buildings with outside air requires sufficiently large ventilation rates driven by a combination of wind and buoyancy effects. Given the complexities of the surrounding urban canopy and the induced flow field, these natural ventilation rates can be highly uncertain and variable. Previous studies have focused either on individual buildings and their surroundings or on very simplified surrounding geometries. This study generalizes characteristics from real urban areas into an idealized canopy, considering ventilation through multiple buildings and window configurations. By simulating wind through this canopy along with airflow through building interiors, we highlight the important effect of the  surrounding geometry on the local flow field and natural ventilation rates. %

\end{boxtext}

\section{Introduction} \label{sec:intro}
The International Energy Agency has projected that if unchecked, the energy demand for building cooling will triple by 2050. This demand would equal China's total electricity consumption in 2018~\citep{dean_future_2018}, pointing to a global need for more sustainable solutions. One alternative is natural cooling, which requires little or no energy by combining natural ventilation with thermal storage. Natural ventilation uses wind and buoyant forces to drive outside air through a building \cite{hunt_fluid_1999}. Often, in the evening or at night, these forces exchange cooler outdoor air with warmer indoor air. This outside air cools building surfaces, which function as thermal storage, absorbing heat and maintaining a comfortable indoor temperature during hotter times of day. Natural cooling systems can replace or augment traditional mechanical cooling, ranging from complex automated systems in larger office buildings (e.g., Stanford's Y2E2~\citep{chen_full-scale_2022}) to opportunely opening windows in residential homes. Although predictive tools for indoor and outdoor temperatures have found widespread implementation in recent decades~\citep{zawadzka_assessment_2021}, characterizing wind and interior airflow across the diversity of urban areas remains challenging~\citep{tong_mapping_2021}.

Building energy simulations often represent natural ventilation with 1-D airflow models~\citep{ramponi_energy_2014}. Specified pressures at building openings force these airflow models. Empirical models derived from wind tunnel data often predict these pressures~\citep{orme_applicable_1999, costola_overview_2009}. However, wind tunnel data generally span limited building and urban geometries~\citep{freire_improvement_2013, costola_overview_2009}, leading most models to account for surrounding buildings with correction factors that are a function of canopy density~\citep{costola_overview_2009}. While these models are widely used~\citep{chiesa_python-based_2019, van_nguyen_new_2018}, they are not broadly accurate~\citep{freire_improvement_2013, ramponi_energy_2014}. Additionally, these models do not consider the coupling between the outdoor and indoor airflow. %

Large-eddy simulations (LESs) offer an alternative approach to predict the coupled outdoor and indoor airflow through specific canopy and building geometries. By resolving the larger turbulence scales that dominate the urban canopy and natural ventilation flows, LESs can accurately predict the airflow and resulting ventilation rates. Previous LESs of coupled indoor and outdoor flow fields often considered idealized building interiors~\citep{van_hooff_accuracy_2017, hwang_large-eddy_2022-1, hirose_indoor_2021} and surroundings~\citep{king_investigating_2017, adachi_numerical_2020}, or were building-specific~\citep{hwang_large-eddy_2023, chen_full-scale_2022}. While these studies have supported the validation of LES for wind-driven ventilation, LES can be used more broadly to understand and quantify natural ventilation with a focus on developing models that represent the surrounding urban canopy's effect. %

The objective of this study is to improve our understanding of the impact of urban canopy geometry on wind-driven natural ventilation and cooling under varying wind conditions. To achieve this objective, we use LESs to predict wind flow through urban canopy and interior building geometries that intend to represent typical characteristics of real urban geometries. We consider urban canopies consisting of quadrants with changing street orientations at two different densities and calculate the ventilation rates through buildings at different locations within the quadrants for different wind directions. The building interiors include four representative ventilation configurations: cross, corner, dual-room, and single-sided. For these four ventilation configurations, we quantify the urban canopy's effect on natural ventilation flow rates by exploring the impacts of canopy density, wind direction, and house location. Through this analysis, we seek to identify the primary factors impacting natural ventilation flow rates in real urban environments. Our results provide a first step towards developing improved models that can efficiently represent the dominant urban canopy effects and support accurate natural ventilation and cooling assessment across large-scale urban areas with diverse geometric characteristics. %

Section~\ref{sec:methods} describes the governing equations, computational setup, and quantification of ventilation rates. Section~\ref{sec:Results} presents the results both as flow visualizations and ventilation rates and discusses the implications for future model development. Section~\ref{sec:Conclusion} summarizes the conclusions.

\section{Description of computational model and quantities of interest} \label{sec:methods}

This section describes the governing equations, computational domain, mesh, boundary conditions, and other forcings. Across 16 simulations, we capture the combined effects of canopy density, wind angle, wind speed, and indoor-outdoor temperature differences on the interior ventilation rates. We measure ventilation rates through four interior sections with different window configurations.

\subsection{Governing Equations and Numerics}

Our simulations use the low-Mach formulation of Cascade Design System's CharLES solver~\citep{cascade_technologies_charles_2022}. 
CharLES is a finite volume solver with an automated body-fitted meshing technique based on 3D-clipped Voronoi diagrams that produces isotropic polyhedral-type cells. Various studies have found good agreement between CharLES and experimental data for natural ventilation and wind pressure loads, including both wind tunnel ~\citep{hwang_large-eddy_2022, ciarlatani_investigation_2023, vargiemezis_predictive_nodate} and full-scale data~\citep{hochschild_comparison_2024, hwang_large-eddy_2023}. Many of these studies included complex urban canopies~\citep{hwang_large-eddy_2023, hochschild_comparison_2024, vargiemezis_predictive_nodate}.

 The low-Mach formulation of CharLES solves the filtered equations for conservation of mass and momentum with the density $\rho$ approximated as the sum of a background density and an isentropic acoustic perturbation. The equations are as follows:

\begin{equation}
\label{eqn:density}
    \frac{\partial \tilde{\rho}}{\partial t} + \frac{\partial \tilde{\rho} \tilde{u}_j}{\partial x_j} = 0,
\end{equation}

\begin{equation}
\label{eqn:NS}
\frac{\partial \tilde{\rho} \tilde{u}_i}{\partial t} + \frac{\partial \tilde{\rho} \tilde{u}_i \tilde{u}_j}{\partial x_j} = - \frac{\partial \tilde{p}}{\partial x_i} + \frac{\partial{\sigma}_{ij}}{\partial x_j},
\end{equation}

\begin{equation}
\label{eqnt:cont}
    \tilde{\rho} = \frac{1}{c^2}(\tilde{p} - p_{\text{ref}}) + \rho_{\text{ref}},
\end{equation}
where $\widetilde{(\cdot)}$ denotes LES filtered quantities, $u_i$ are the velocity components, $p$ is the pressure, ${\sigma_{ij}}$ is the stress tensor, $c$ is the speed of sound, $p_\text{ref}$ is the reference pressure, and $\rho_\text{ref}$ is the reference density. The unresolved portion of the stress tensor is modeled with the Vreman subgrid model~\citep{vreman_eddy-viscosity_2004}.

The equations are discretized using a second-order backward difference scheme in time and a second-order central discretization in space. Defining a finite speed of sound results in a lower condition number for the pressure system of equations, which is a Helmholtz equation instead of the Poisson equation that arises in fully incompressible formulations. In the zero Mach number limit, the system will discretely recover an incompressible formulation. Additional insights into the derivation of the Helmholtz system can be found in \citep{ambo_aerodynamic_2020}.

\subsection{Computational Domain and Grid} \label{ssec:domain}
This section begins by describing the urban canopy layouts before detailing the building interiors and computational grid.

\subsubsection{Urban Canopy Layout}
\label{sssec:CanopyLayout}

\begin{figure}[h!]
    \centering
    \includegraphics[width=0.6\textwidth]{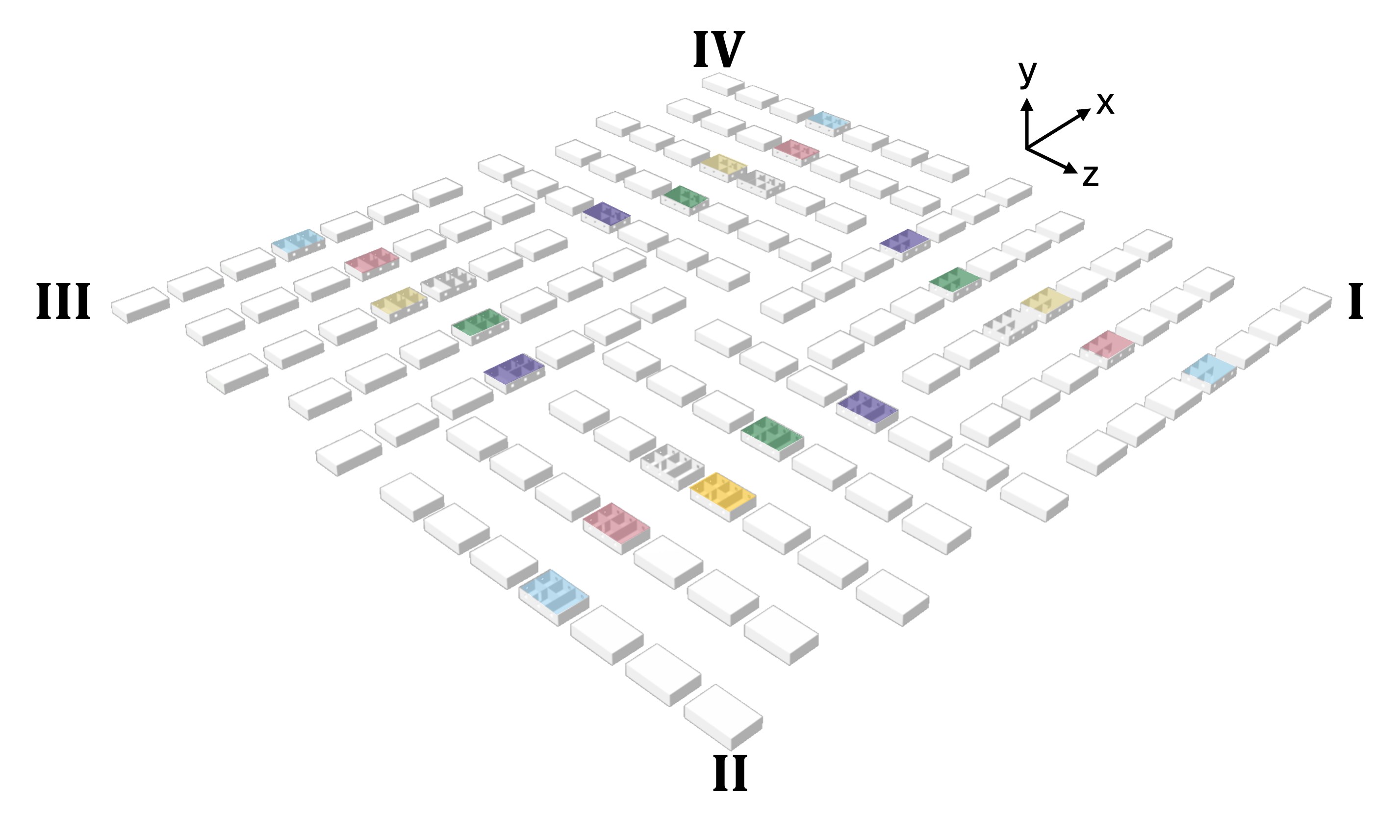}
    \caption{House placement within the simulation domain. Colors mark interiors across each quadrant.}
    \label{fig:House Locations}
\end{figure}

Representing the characteristics of real urban canopies is challenging ~\citep{lu_novel_2023, adolphe_simplified_2001, xue_quantifying_2022, biljecki_global_2022}. In this study, we consider an idealized urban area that aims to represent a few key characteristics of real urban canopies. Figure~\ref{fig:House Locations} shows the locations of houses across the computational domain. In total, the computational domain includes 140 buildings distributed over four quadrants. Each quadrant contains an identical group of 35 houses that form a five-by-seven staggered grid, with houses staggered in the seven-house direction (along side yards) and aligned in the five-house direction (along streets). Each quadrant is rotated by $90^o$ to define the full domain. The borders between quadrants are equally-sized within the domain and across the boundaries. 

Simulating four quadrants presents a couple of advantages. First, alternating orientations prevent infinitely repeating urban streets, which are not representative of real urban geometries~\citep{lu_novel_2023}. Second, the borders between the four quadrants act as discontinuities, similar to those in real urban layouts, providing greater geometric diversity. In this four-quadrant domain, we run wind directions parallel and diagonal with respect to the urban grid. Across these two simulations, the identical group of 35 houses in each quadrant experiences eight wind directions, as outlined in Table \ref{table:WindAngles}. Parallel wind travels along streets in the \ang{0} and \ang{180} quadrants and along side yards in the \ang{90} and \ang{135} quadrants. Diagonal wind is diagonal to both streets and side yards in all quadrants.

\begin{table}[h]
    \centering
    \begin{tabular}{lcccc}
    \hline
    \textbf{} & \textbf{I} & \textbf{II} & \textbf{III} & \textbf{VII} \\ \hline
    Parallel & \ang{0} & \ang{270} & \ang{180} & \ang{90} \\ 
    Diagonal & \ang{45} & \ang{315} & \ang{225} & \ang{135} \\ \hline
    \end{tabular}
    \caption{Wind angles experienced by each quadrant under wind parallel and diagonal to urban grid.}
    \label{table:WindAngles}
\end{table}

\begin{figure}[h!]
    \centering
    \includegraphics[width=0.8\textwidth]{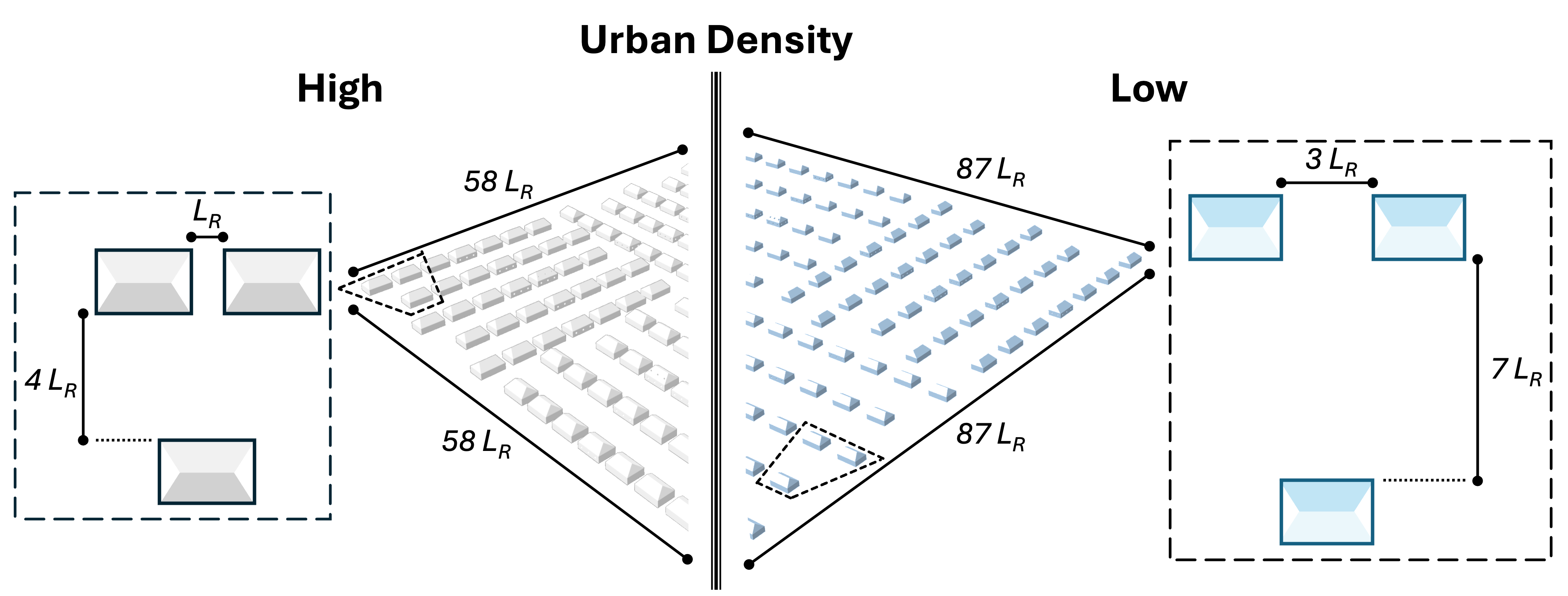}
    \caption{Canopy horizontal dimensions with $L_R = 4 \text{ m}$. }
    \label{fig:Canopy Plan}
\end{figure}

We consider two domains representing different canopy densities. The house centers are $50\%$ further apart in the low density domain, but the house dimensions remain the same. Figure~\ref{fig:Canopy Plan} shows the horizontal dimensions of both canopies, with the width of the rooms $L_R = 4\text{ m}$ as the reference length.
The heights of both domains ($H$) are one-third of their lengths.

Table \ref{table:frontalArea} summarizes the corresponding frontal area fractions ($\lambda_f$) of the urban canopies. The frontal area fraction is commonly used to parameterize an urban canopy's influence on air flow~\citep{wang_wind_2024, lu_novel_2023}. $\lambda_f$ describes the proportion of the urban area occupied by the urban canopy when looking in the direction of the mean wind. Table \ref{table:frontalArea} presents $\lambda_f$ along the aligned, staggered and diagonal axes.
\begin{table}[h]
    \centering
    \begin{tabular}{lccc}
    \hline
    \textbf{Canopy} & \textbf{Aligned} & \textbf{Staggered} & \textbf{Diagonal} \\ \hline
    High Density & 1/3 & 3/4 & 13/24 \\ 
    Low Density  & 2/7 & 1/2 & 11/28 \\ \hline
    \end{tabular}
    \caption{Frontal area fractions ($\lambda_f$) for the urban canopy below $H_{R}$.}
    \label{table:frontalArea}
\end{table}

\subsubsection{Building Details}

\begin{figure}[h!]
    \centering
    \includegraphics[width=0.6\textwidth]{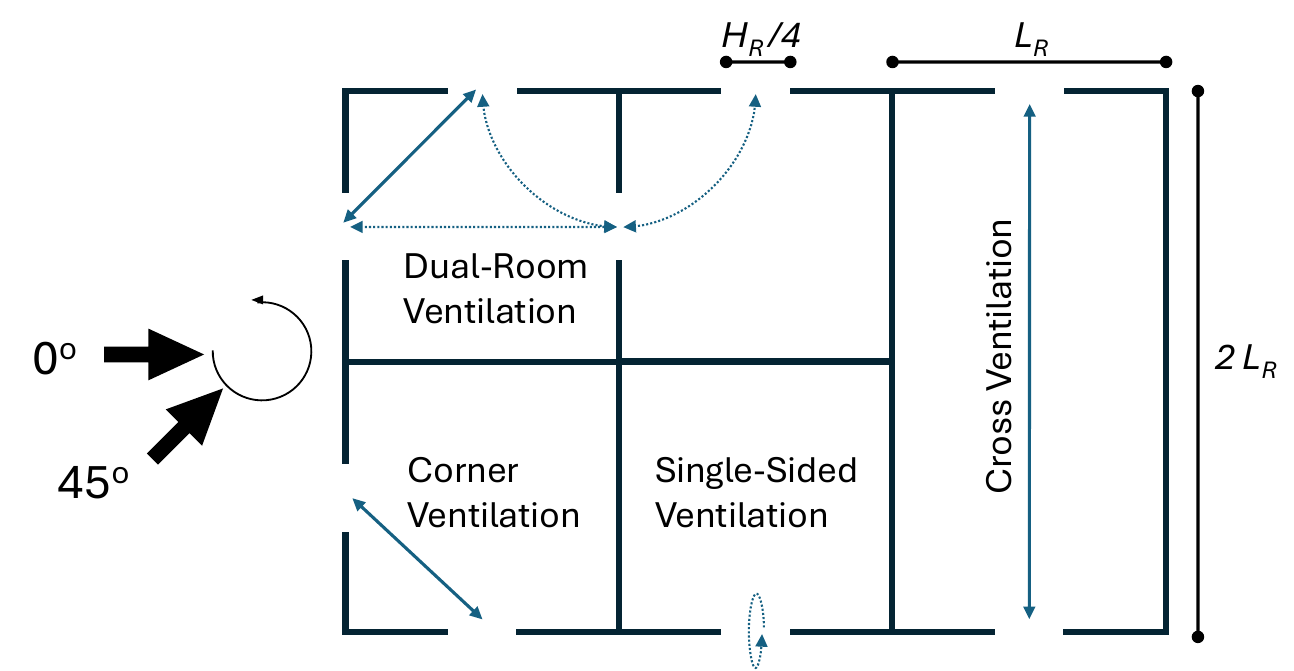}
    \caption{Floor plan of simulated interiors. Arrows indicate possible ventilation pathlines, with solid arrows showing ventilation axes. }
    \label{fig:Floor Plan}
\end{figure}

The overall dimensions of each house are 3 $L_R$ by 2 $L_R$, with $L_R = 4 \text{ m}$  being the width of interior rooms. The house has a hip roof with a minimum height equal to the interior room height $H_R = 3\text{ m}$ and a maximum height $H_H = 2H_{R}$. The roof features a single ridge extending 2 $L_{R}$ over the house center in the aligned direction. Within each quadrant of 35 houses, six have simulated interiors. While one of the six houses includes skylights, this study focuses on the five houses without skylights, all with identical interiors. 

Figure~\ref{fig:Floor Plan} shows the floor plan. The interiors include four separate sections that do not exchange airflow. Three of the sections are individual rooms, while the fourth is two connected rooms. Most rooms are square; the cross-ventilated room is the exception, being $L_{R}$ wide and 2 $L_{R}$ long. 

The window placement gives each section a unique ventilation type, as indicated in Figure~\ref{fig:Floor Plan}. All windows are $H_{R}/4$ by $H_{R}/4$. The two rooms in the dual-ventilated section are connected by an open door frame $H_{R}/4$ wide by 3 $H_{R}/4$ tall. The arrows in Figure~\ref{fig:Floor Plan} trace possible flow paths for each ventilation type. The straight path between two windows will be referred to as the ventilation axes in the remainder of this paper.

\subsubsection{Computational Grid and Time Step}

\begin{figure}[h!]
    \centering
    \includegraphics[width=0.55\textwidth]{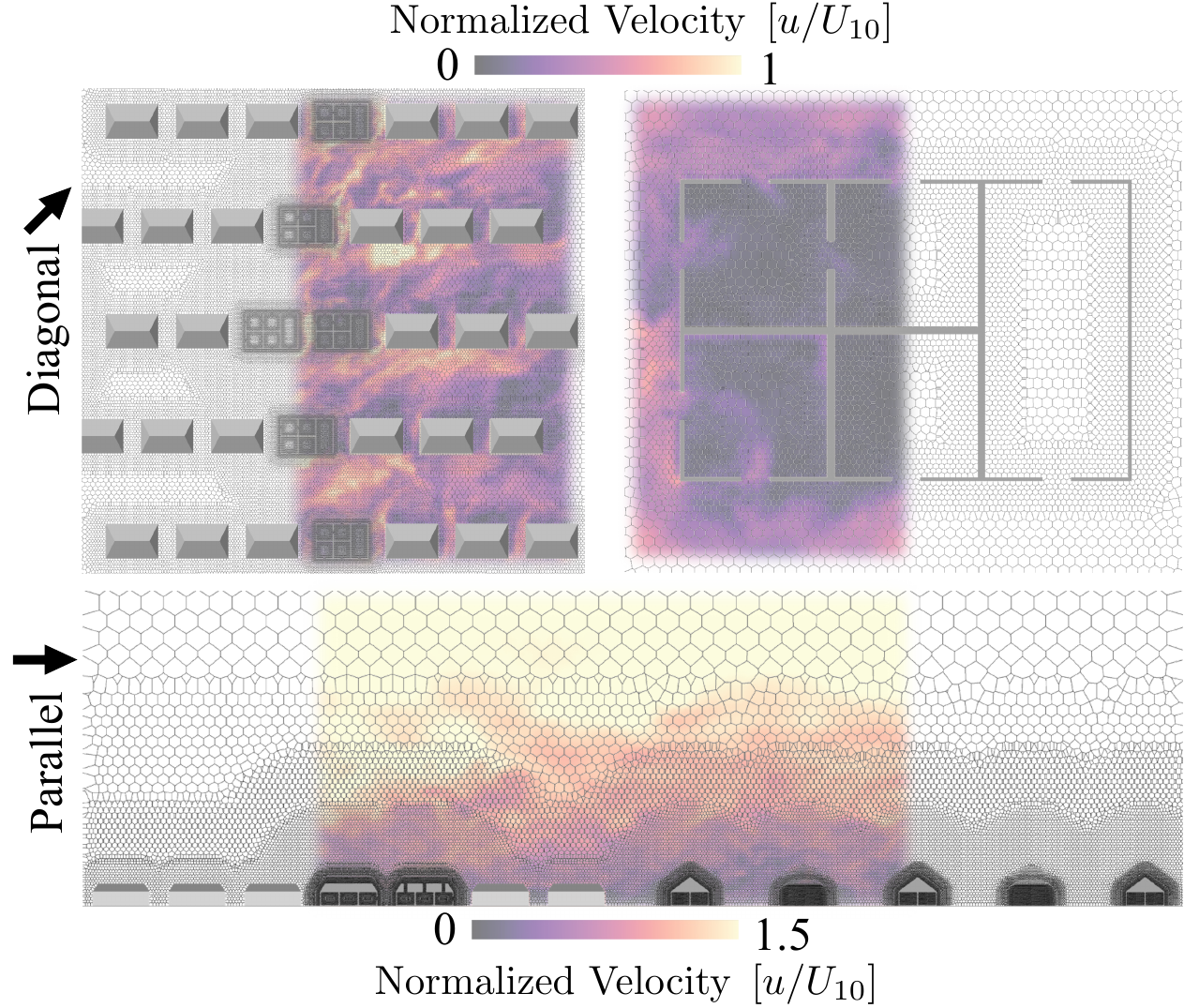}
    \caption{Turbulent flow snapshots from parallel and diagonal wind layered over high density canopy mesh. Plan views at window center (top) and profile through two quadrant centers (bottom).}
    \label{fig:Mesh}
\end{figure}

Figure~\ref{fig:Mesh} shows cross-sections of the mesh created by CharLES's body-fitted mesh generation based on 3D-clipped Voronoi diagrams. Each window has five cells across the height and width~\citep{hwang_large-eddy_2022-1}, for a mesh resolution of $H_R/20$. For houses with simulated interiors, this mesh resolution extends for six elements around interior and exterior walls The mesh then coarsens to $H_R$/10 for a distance of $L_R/2$ before coarsening again to $H_R/5$, or 10 cells per canopy height. This resolution extends across the adjoining streets to resolve local canopy effects. This mesh resolution also continues through all side yards and within $L_R$ of houses without simulated interiors. Away from these refinement zones, the mesh size in the canopy is $2H_R/5$. Moving up from the canopy, the mesh size doubles every five elements, eventually coarsening to $16H_R/5$. For the high and low density canopies, this mesh resolution leads to approximately 9 and 14 million elements, respectively. To check mesh sensitivity, we halved the cell size and simulated the high density canopy with approximately 53 million elements. This mesh refinement did not significantly change the results, with an average discrepancy of 6\%. Appendix A (included in supplementary materials) shows detailed results for this mesh refinement study. For all simulations, the timestep is adjusted to keep the Courant–Friedrichs–Lewy (CFL) condition less than one during the majority of each simulation: 0.02 seconds for $U_{10} = 2 \text{ m/s}$ and 0.01 seconds for $U_{10} = 4 \text{ m/s}$.

\subsection{Boundary Conditions and Periodic Forcing} \label{ssec:BoundaryConditions}

\subsubsection{Wind and Temperature Conditions}
\label{sssec:UandTConditions}
The wind speed and temperature conditions used in this study were derived using typical-year weather data from California's 16 climate zones~\citep{california_energy_commission_ca_2022} in combination with a building energy model (BEM). We focus on two cooling scenarios: one in the evening and one at night. For the evening scenario, natural cooling initiates after 4:00 pm as soon as the outdoor temperature drops below the indoor temperature. For the nighttime scenario, natural cooling initiates after 11:00 pm if the outdoor temperature is below the indoor temperature. At each ventilation time, we compile wind speeds and indoor-outdoor temperature differences observed in a typical year. Based on this analysis, we selected two typical wind and temperature scenarios: 10 m height wind speeds of 2 and 4 m/s and indoor-outdoor temperature differences of 0 and 5 \SI{}{\degreeCelsius}. 

We performed simulations for the four possible combinations of these conditions, assuming neutral conditions for the outdoor wind flow. This assumption is informed by our focus on evening and night-time ventilation scenarios, but even for a non-neutral outdoor flow, temperature effects can be expected to be relatively small since most ventilation is driven by flow in the isolated roughness regime ~\citep{oke_street_1988,xue_impact_2024, xie_impact_2007}. Furthermore, the results presented in this paper only consider ventilation through window openings at a single height, where buoyant effects are small. As a result, we found that the wind speed and indoor-outdoor temperature difference had negligible effects on the non-dimensional velocity and ventilation rates. We therefore average over the set of four simulations for all results in this paper, with each simulation given equal weighting.

\subsubsection{Boundary Conditions}
We represent the top boundary as a slip condition, the horizontal boundary conditions as periodic, and the ground and building surfaces as smooth walls (using a standard algebraic wall model). We found the choice of the wall model (e.g., smooth or rough wall) has a negligible impact on the predicted ventilation rates since the flow is dominated by large-scale sharp-edge separation and wakes. We initialize all surfaces and air volumes with the indoor-outdoor temperature differences from the BEM (see Section~\ref{sssec:UandTConditions}). The surface temperatures remain fixed while the air volumes mix during the simulation spin-up to a statistically steady-state flow and temperature field.

\subsubsection{Periodic Wind Forcing}
\label{sssec:Periodic Wind Forcing}
\begin{figure}[h!]
    \centering
    \includegraphics[width=0.9\textwidth]{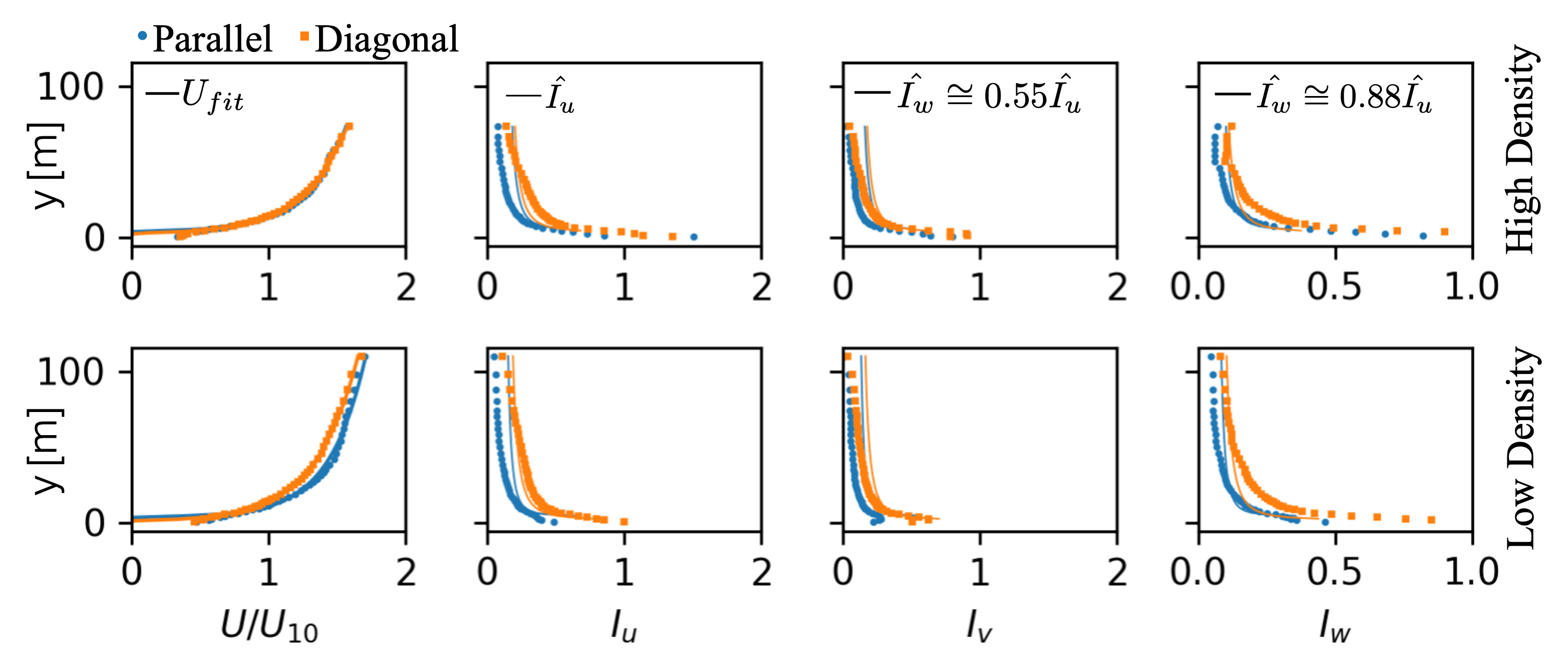}
    \caption{Average profiles of mean velocity ($U$) and turbulence intensities ($I$), with $x$ aligned with the mean flow. Lines in the top row show fitted ABL profiles ($U_{fit}$) while subsequent lines plot the empirical turbulence intensities using $\hat{I_u} \cong 1/\log \left((y\!-\!d)/y_0\right)$~\citep{holmes_wind_2018}. Profiles span the domain height.}
    \label{fig:ABL Profiles}
\end{figure}

A momentum source drives the flow to maintain a constant spatially averaged velocity in the top 10\% of the domain. We assume the velocity near the domain height ($H$) reflects an ABL driven by geostrophic wind over a large urban area. Below $H$, local canopy characteristics govern the boundary layer~\citep{stull_introduction_2003, letzel_high_2008}.  While this local boundary layer depends on the choice of $H$, larger $H$'s have increasingly smaller effects due to the logarithmic ABL profile (see \ref{eqn:ABL}). To set the velocity at $H$, we assume a suburban terrain with bulk roughness $y_0 = 0.3$~\citep{aboshosha_abl_2015, stull_introduction_2003} and disregard the 6 m canopy height, setting the displacement height $d=0$ for simplicity. We then prescribe a 10 m height wind speed ($U_{10}$), and calculate the associated velocity at $H$ using the mean ABL profile:
\begin{equation}
    \label{eqn:ABL}
    U(y) = \frac{u_\ast}{\kappa}\ln{\left(\frac{y - d}{y_{0}}\right)}
\end{equation}
We choose the friction velocity ($u_\ast$) to yield the desired $U_{10}$ and set the von Kármán constant to $\kappa = 0.41$. The simulations initialize from a laminar flow field. Then, as the flow passes through the periodic domain, the boundary layer adjusts to the urban canopy yielding a well-developed turbulent surface layer~\citep{philips_large-eddy_2013, grylls_steady-state_2020}. Figure~\ref{fig:ABL Profiles} shows profiles of mean velocity and turbulence intensities averaged i) along horizontal domain extents, ii) over runtime, and iii) across simulations of the four $U_{10}$ and $T_{in} - T_{out}$ combinations. Table \ref{table:ABLFits} lists $u_\ast$, $y_0$, and $d$ fit to the mean velocity profiles. These fitted $y_0$ and $d$ show the large effects of canopy density and wind alignment on the resulting boundary layers, with $y_0$'s both above and below the initial $y_0 = 0.3$ estimate.

\begin{table}[h!]
    \centering
    \begin{tabular}{lcc|cc}
        \hline
        \textbf{ABL Parameter} & \multicolumn{2}{c|}{\textbf{High Density}} & \multicolumn{2}{c}{\textbf{Low Density}} \\
        \cline{2-5}
        & Parallel & Diagonal & Parallel & Diagonal \\
        \hline
        \textbf{$u_\ast / U_{10}$} & 0.14 & 0.12 & 0.13 & 0.09 \\
        \textbf{$y_0$} & 0.63 & 0.31 & 0.56 & 0.05 \\
        \textbf{$d$} & 2.59 & 3.17 & 1.52 & 4.22 \\
        \hline
    \end{tabular}
    \caption{ABL parameters fit to mean velocities above canopy.}
    \label{table:ABLFits}
\end{table}

\subsection{Summary of LES runs}

In total, we perform a set of 16 simulations, reflecting all possible combinations of four parameters: canopy density, wind direction, wind speed, and indoor-outdoor temperature difference. Table \ref{table:LESparams} summarizes these parameters and their associated values. As mentioned in Section~\ref{sssec:UandTConditions}, the results presented in this paper were obtained by averaging across different wind speeds and temperature differences, producing four data sets for any given combination of canopy density and wind alignment. Due to the varying quadrant orientation within the domain, the two wind directions (i.e. `Parallel' and `Diagonal') provide data for ventilation rates at eight different wind angles (see Section~\ref{sssec:CanopyLayout}). 

\begin{table}[h]
    \centering
    \begin{tabular}{lcc}
    \hline
    \textbf{Parameter} & \textbf{Option 1} & \textbf{Option 2} \\ \hline
    Canopy Density & High & Low \\
    Wind Direction & Parallel & Diagonal \\
    $U_{10}$ & 2 m/s & 4 m/s \\
    $T_{in} - T_{out}$ & 0 \SI{}{\degreeCelsius} & 5 \SI{}{\degreeCelsius}\\ \hline
    \end{tabular}
    \caption{Parameters varied across LES simulations.}
    \label{table:LESparams}
\end{table}

The simulations were spun-up over at least four flow passes and subsequently data was collected over at least eighteen flow passes. %
We determined that these spin-up and data collection times were sufficient based on ventilation rate and flow field convergence. On 4 AMD Epyc 7502 nodes, totaling 128 cores, running a flow-through period for the low-density canopy takes approximately eight hours.

\subsection{Quantification of Ventilation Rates}
From the LES time series for the velocity field we can calculate the instantaneous ventilation rate ($Q(t)$) as:
\begin{equation}
     \label{eqn:Qins}
     Q(t) = \frac{1}{2}\sum_{i=1}^{N_w} \int_{A_i} \left| \mathbf{u}(t) \cdot \mathbf{n}_i \right| \, dA,
\end{equation}
where, for a given interior space with $N_w$ windows, $A_i$ is the area of each window and $n_i$ is the associated normal vector~\citep{hwang_large-eddy_2022-1}. $Q$ accounts for turbulent exchange at individual windows by using the absolute values of velocities. The factor of $1/2$ prevents double-counting flow into and out of the interior.
A time-averaged, non-dimensional ventilation rate is then obtained from 
\begin{equation}
     \label{eqn:Q}
     Q_n = \frac{\sum_{i=1}^{N_t}Q(t)}{U_{10} A}, 
\end{equation}
where $N_t$ is the number of time steps, and A is the total window area divided by 2.

To analyze and interpret the variability in the calculated ventilation rates, we use the median and the interquartile range (IQR). The IQR is the distance between the $25^{th}$ and $75^{th}$ percentiles ($\text{IQR} = p75 - p25$), or equivalently the range spanned by the center 50\% of results when ordered from smallest to largest. We chose the median and IQR because these estimators are interpretable and robust to outliers. To understand the effects of different parameters on ventilation rates, we condition the median and IQR estimates on different parameters. Conditioning on more parameters isolates increasingly specific ventilation scenarios, generally decreasing the IQR. Because this conditioning also reduces the sample size, we report the conditional sample-normalized IQR, defined as

\begin{equation}
\label{eqn::IQRn}
\text{IQR}_n(Q_n \mid P_{i:j}) = \text{IQR}(Q_n \mid P_{i:j}) \cdot\Biggr[ \frac{\text{IQR}(Q_n)}{\mathbb{E}_{\text{boot}}[\text{IQR}_n(Q_n)]} \Biggr],
\end{equation}
where $P$ are the parameters specifying the canopy, as reported in Table \ref{table:LESparams}. Dividing $\text{IQR}(Q_n)$ by $\mathbb{E}_{\text{boot}}[\text{IQR}_n(Q_n)]$, the bootstrapped estimate of the IQR at that sample size, gives a sample size correction. This correction compares the effect of conditioning on a given parameter with that of randomly selecting a subset of ventilation rates.

\section{Results and Discussion} \label{sec:Results}
In this section, we first present visualizations of the mean flow field at two spatial scales: across the full domain and around individual buildings. Next, we quantify ventilation rates and their sensitivity to the parameters listed in Table \ref{table:LESparams}. To explain the ventilation rate sensitivities, we connect ventilation trends with the observed canopy flow characteristics.

\subsection{Flow Field Visualization} 
\label{ssec:Flow Field}

This section first describes qualitatively how the canopy-scale flow depends on (i) canopy density (high vs low), (ii) building alignment (aligned vs staggered), and (iii) wind direction (parallel vs diagonal). Next, it considers the more local flow field around individual buildings to elucidate how canopy flow effects interact with flow through building interiors. All flow visualizations are at window-center height ($H/2$) and show the normalized velocity ($U/{U_{10}}$) %
averaged across simulations for the two wind speeds and two indoor-outdoor temperature differences (see Section~\ref{sssec:UandTConditions}). %

\subsubsection{Domain-Scale Canopy Flow} \label{sssec:Canopy Flow}

\begin{figure}[hb]
    \centering
    \includegraphics[width=0.8\textwidth]{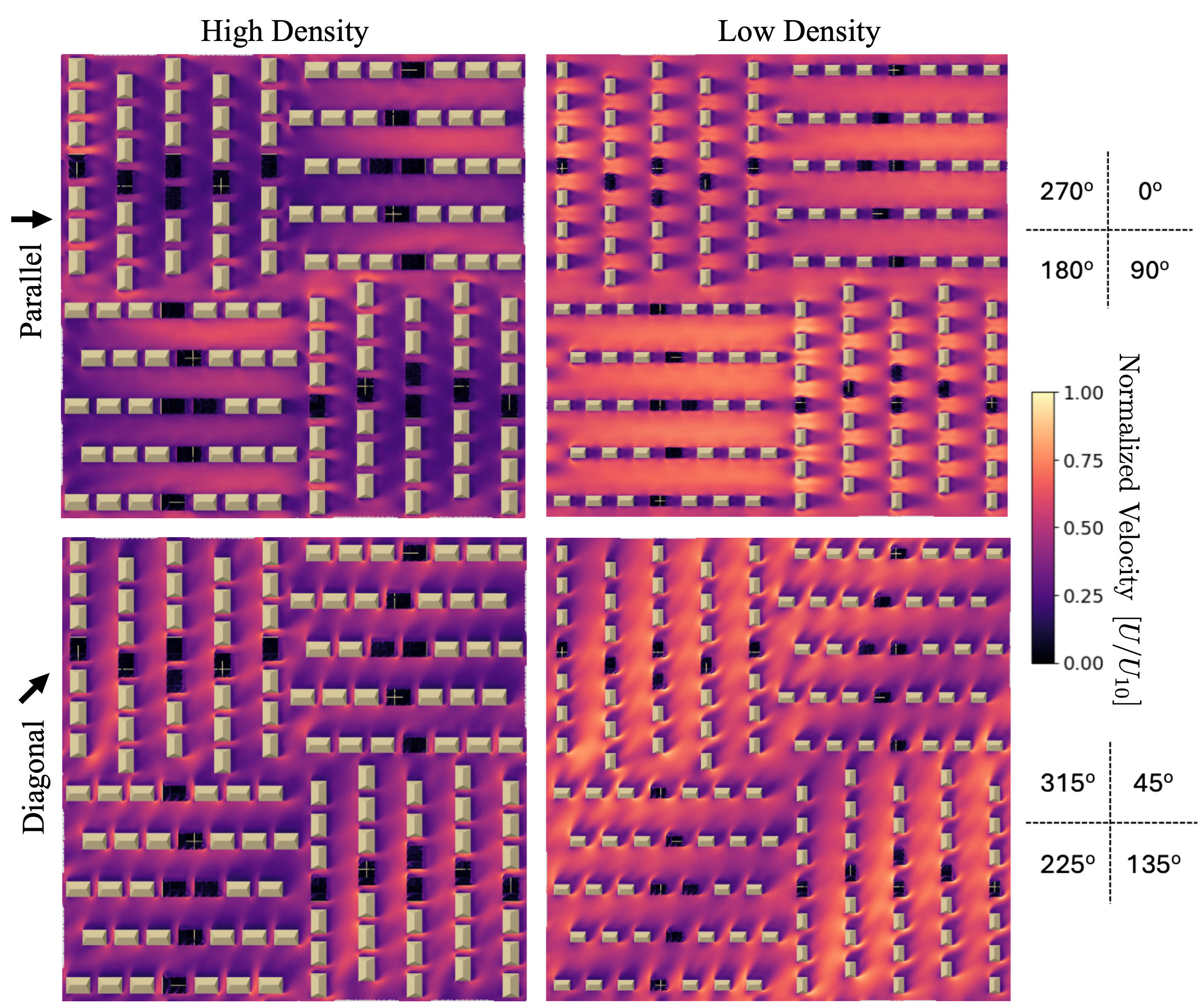}
    \caption{Flow fields at window-center height ($H/2$). Keys on the right side show the wind angle (relative to the house) associated with each quadrant in the simulation domain.}
    \label{fig:Flow Fields}
\end{figure}

Figure~\ref{fig:Flow Fields} shows contour plots of the mean velocity magnitude for the two canopy densities and for parallel and diagonal flow. The plots reveal significant large-scale effects of urban canopy density, building alignment, and wind direction on the in-canopy flow. 

As expected, the lower density domains support a faster in-canopy flow, consistent with the smaller $\lambda_f$ (see Table \ref{table:frontalArea}). This trend also appears between the aligned and staggered quadrants under parallel flow. The aligned quadrants, which have a smaller $\lambda_f$, support an overall faster in-canopy flow with the building wakes confined to the smaller side yards. The staggered quadrants cause building wakes persist through much of the canopy. At the transition between quadrants with different alignments, the overall higher (or lower) velocities from the aligned (or staggered) quadrant persist into the downstream quadrant. Overall, these results show that areas of high $\lambda_f$ cause reductions in canopy flow velocity which may persist some distance downstream. 

\begin{figure}[htb]
    \centering
    \includegraphics[width=0.4\textwidth]{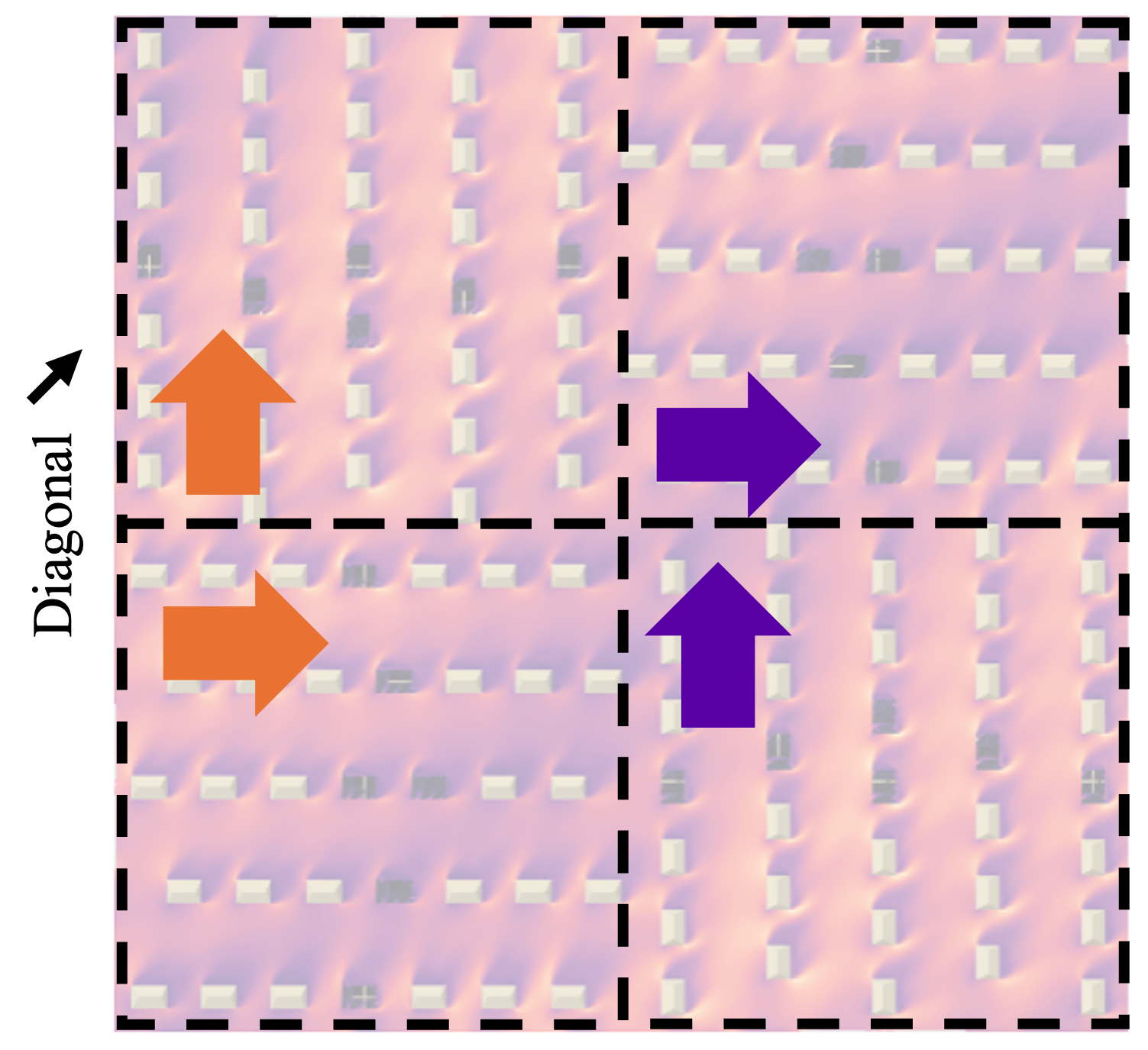}
    \caption{Arrows show diagonal wind projected along streets. The orange arrows show canopy flow divergence (arrows pointing apart), whereas the purple arrows show canopy flow convergence (arrows pointing together). This coloring creates a similar pattern to the diagonal flow field.} 
    \label{fig:Diagonal Schematic}
\end{figure}

The diagonal flow cases in Figure~\ref{fig:Flow Fields} also reveal domain-scale variability in wind speed, even though $\lambda_f$ is constant across quadrants. This variation stems from the anisotropic nature of the drag exerted by a quadrant on the flow, i.e., the drag is lower in the direction of the aligned streets. Figure~\ref{fig:Diagonal Schematic} schematically represents the deflection of canopy flow along the streets, as indicated by the arrows in each quadrant. This deflection directs the flow in the canopy to either diverge (orange arrows) or converge (purple arrows), and this diverging/converging pattern sets up a secondary flow pattern. Where the canopy flow diverges, faster flow from above the canopy descends into the canopy, resulting in a higher in-canopy velocity; where the canopy flow converges, slower flow from within the canopy rises and the in-canopy velocity remains lower. The vertical velocities form larger secondary vortices that align with the mean flow direction. Other work has observed these secondary flows over alternating areas of high and low roughness~\cite{ma_inter-scale_2025, anderson_numerical_2015}. In summary, the domain-scale flow patterns reveal that both $\lambda_f$ and anisotropy in the urban canopy drag can cause large-scale variations in wind speed that will impact interior flow and ventilation rates.

\subsubsection{Local Flow Patterns}

\begin{figure}[h!]
    \centering
    \includegraphics[width=0.7\textwidth]{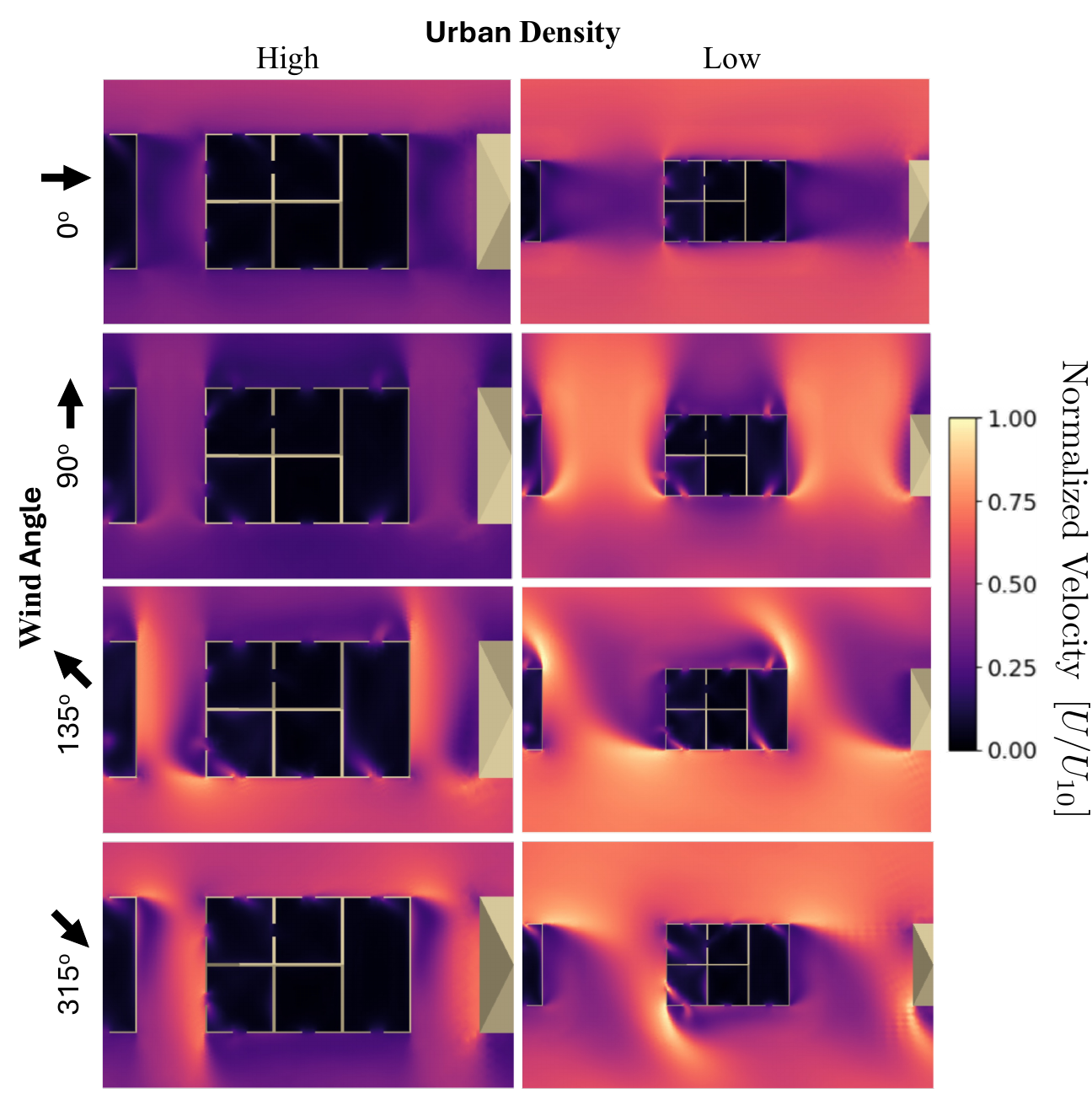}
    \caption{Mean velocity magnitude at window-center height for four wind angles. Interiors are located in the center of their respective quadrant.}
    \label{fig:Local Flow Fields}
\end{figure}

Figure~\ref{fig:Local Flow Fields} shows the time-averaged velocity magnitude around and inside a single house for four wind angles. Lower in-canopy wind speeds in the high-density canopy result in lower flow velocities through window openings. In addition, the interference effects due to surrounding buildings are more significant in the high-density canopy, producing considerably different wind speeds and directions near openings than what would be observed for an isolated building. These changes affect the local momentum and pressure near the openings, which in turn impact ventilation flow. 

Considering the \ang{0} case, the dual- and corner-ventilated rooms have upstream openings that would support significant ventilation flow in the absence of surrounding buildings. However, in the canopy layouts considered in this study, these openings are in the wake of the upstream building and the flow through the openings is small. This effect is larger in the high-density case, which has a skimming flow regime, than in the low-density case, which has an interference flow regime, following the common classification~\citep{oke_street_1988, hussain_wind_1980}. A similar effect is observed for the corner- and cross-ventilated rooms under the \ang{90} wind direction. In the high-density canopy, flow rates through window openings are lower, while in the low-density canopy, flow rates are higher due to the larger distance to upstream buildings and wider side yards.

For the non-parallel flow directions, we similarly observe important interference effects. In particular, interference effects can significantly alter the local flow direction. For example, in the absence of interference effects, the flow through the corner-ventilated room would be similarly aligned with the ventilation axis for the \ang{135} and \ang{315} wind directions. However, in the high-density canopy, the surrounding buildings induce changes in the outdoor flow field that yield higher flow rates for the \ang{135} wind direction than for the \ang{315} wind direction. In the low-density canopy, this effect is reversed. This example demonstrates that the observed changes in the flow field compared to the flow around an isolated building (see for example~\cite{hwang_large-eddy_2022}) strongly depend on the urban canopy layout. We expect a better representation of these interference effects will be key in improving natural ventilation predictions.

\subsection{Ventilation Rates}

This section presents the ventilation rates and their sensitivity to canopy density, wind direction, and house location. We address these parameters from easier to more difficult to characterize. We first discuss the effect of canopy density, which can be described with bulk parameters like $\lambda_f$ and leads to canopy-wide changes in wind speed, as shown in Figure~\ref{fig:Flow Fields}. We then address wind angle and house location, which are governed by local canopy flow features.  

\subsubsection{Ventilation Rates by Room Type and Overall Sensitivity to Canopy Density}

\begin{figure}[htb]
    \centering
    \includegraphics[width=0.8\textwidth]{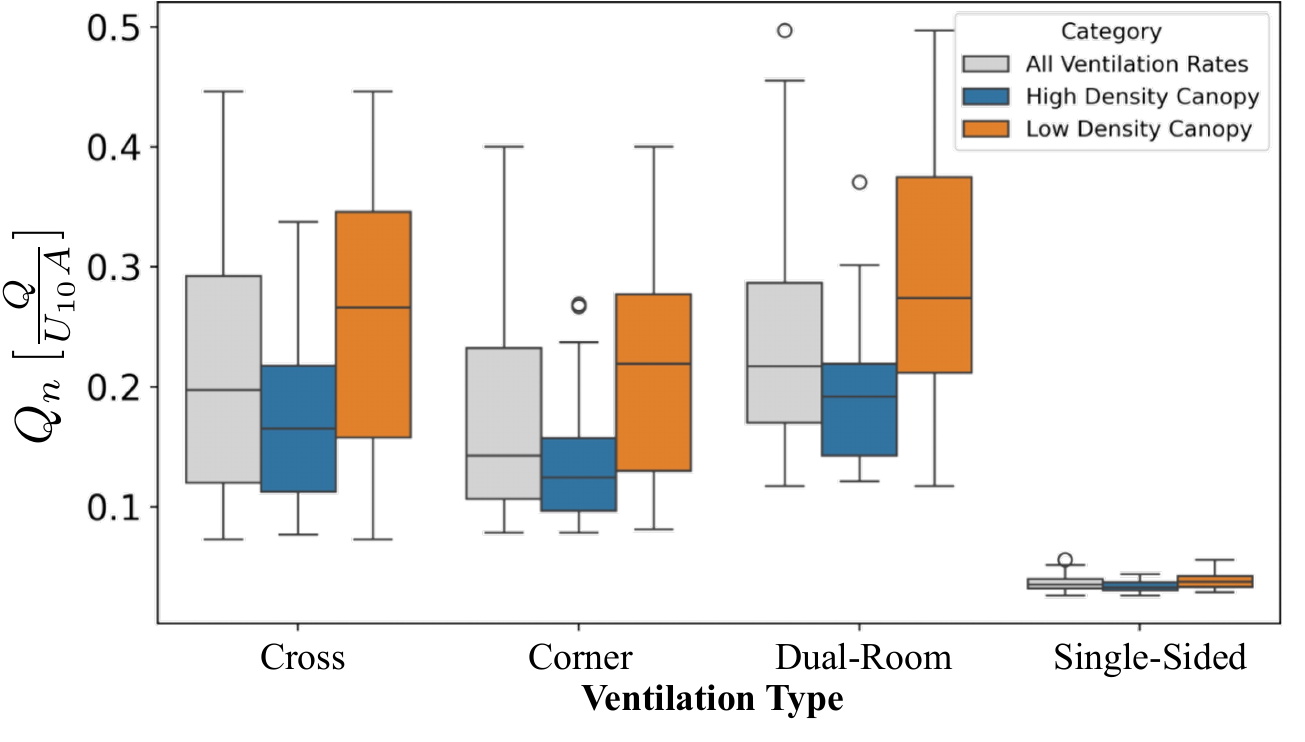}
    \caption{Average normalized ventilation through the four room types. For each room, the first box plot shows ventilation rates across all cases while the second two are conditioned on canopy density. Boxes span from the lowest 25\textsuperscript{th} percentile to the upper 75\textsuperscript{th} percentile, with the box length giving the IQR. Whiskers extend 1.5 times the IQR beyond the nearest quartile. Points show ventilation rates beyond this range.}
    \label{fig:Rooms}
\end{figure}

Figure~\ref{fig:Rooms} shows the ventilation rate distribution by room type. For each room, the first box plot includes the ventilation rates across all LESs, spanning the parameters outlined in Table \ref{table:LESparams}. These plots show that the dual-room configuration experiences the largest median $Q_n$ at 0.22, followed by the cross (10\% less), corner (35\% less), and single-sided (85\% less) ventilation types. The IQR for dual-room ventilation is about 50\% of the median, while the IQR for corner and cross ventilation are about 85\% and 90\% of their respective medians. The average IQR across these three rooms is 75\% higher than the difference in ventilation rates between them. This result indicates that, provided there are multiple windows, the ventilation configuration has less impact than the parameters influencing the external flow field.

To quantify the effect of canopy density, the second and third box plots in Figure~\ref{fig:Rooms} show each room's ventilation rate for the high and low canopy densities, respectively. On average, the higher density canopy gives 15\% lower ventilation rates and 40\% lower IQRs. The lower canopy density gives 30\% higher ventilation rates with 20\% higher IQRs. These numbers confirm that a building in a higher density canopy will experience stronger interference effects, where the surrounding canopy geometry starts to dominate the flow and sensitivity to wind angle is reduced. This reduced sensitivity to wind angle will be further explored in Section~\ref{sssec:Wind Angle}. 

\begin{figure}[htbp]
    \centering
    \includegraphics[width=0.8\textwidth]{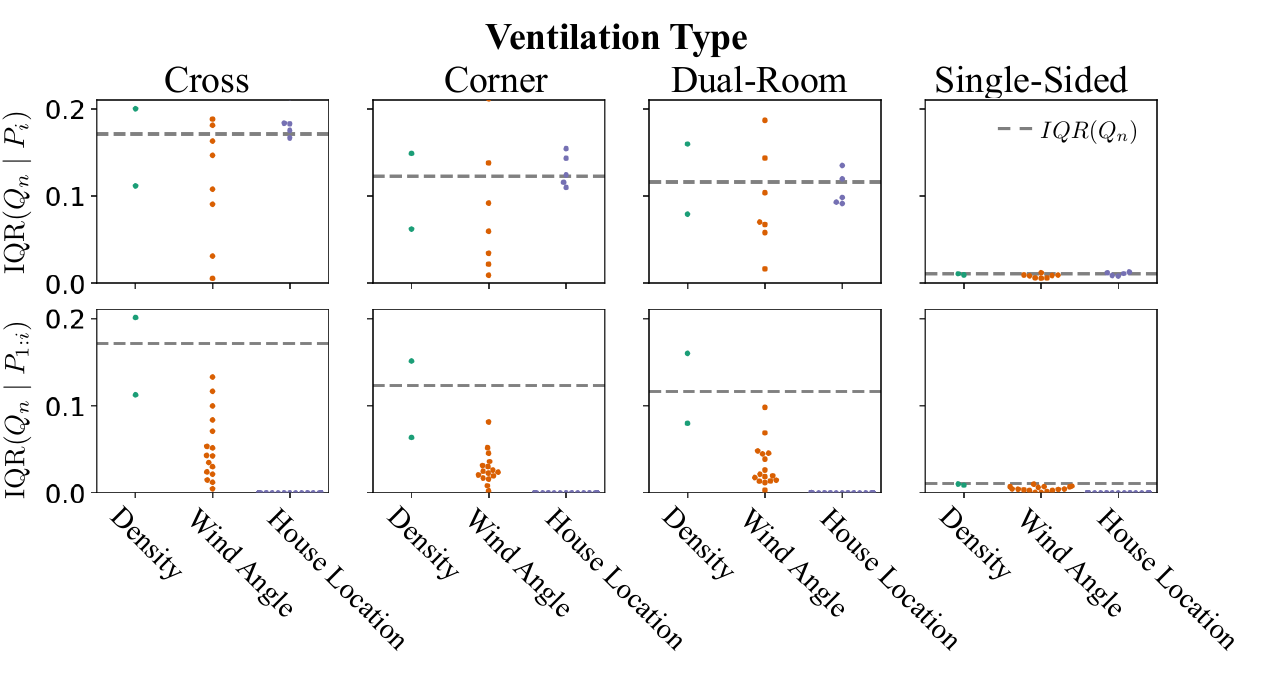}
    \caption{Horizontal lines show the IQR of all $Q_n$ measured in each room, while points show the $\text{IQR}_n$ of $Q_n$ conditioned on a parameter subset. In the top row, each $\text{IQR}_n$ conditions on a single fixed parameter (\( P_i \)). For example, fixing house location results in 5 subsets. In the bottom row, each $\text{IQR}_n$ conditions on both the given parameter and all preceding parameters on the x-axis (\( P_{1:i} \)), meaning more parameters are progressively fixed. For example, fixing wind angle and density results in 16 subsets.} 
    \label{fig:IQR}
\end{figure}
To further illustrate the effects of the different parameters, Figure~\ref{fig:IQR} plots the sample-normalized IQRs, or $\text{IQR}_n$s, for the canopy density, wind angle, and house location. The top row provides $\text{IQR}_n$s conditioned on values from a single parameter, showing that the wind angle, followed by the canopy density, has the greatest potential to reduce $\text{IQR}_n$ in aggregate. The bottom row of Figure~\ref{fig:IQR} quantifies the joint effect of multiple parameters in aggregate by progressively conditioning $\text{IQR}_n$s on more parameters. For multi-window ventilation types, conditioning on both density and wind angle provides a significant decrease in $\text{IQR}_n$s, showing that natural ventilation models should represent the combined effect of these two parameters in order to make accurate predictions. %

\subsubsection{Sensitivity of Ventilation Rate to Wind Angle} \label{sssec:Wind Angle}

\begin{figure}[htbp]
    \centering
    \includegraphics[width=0.6\textwidth]{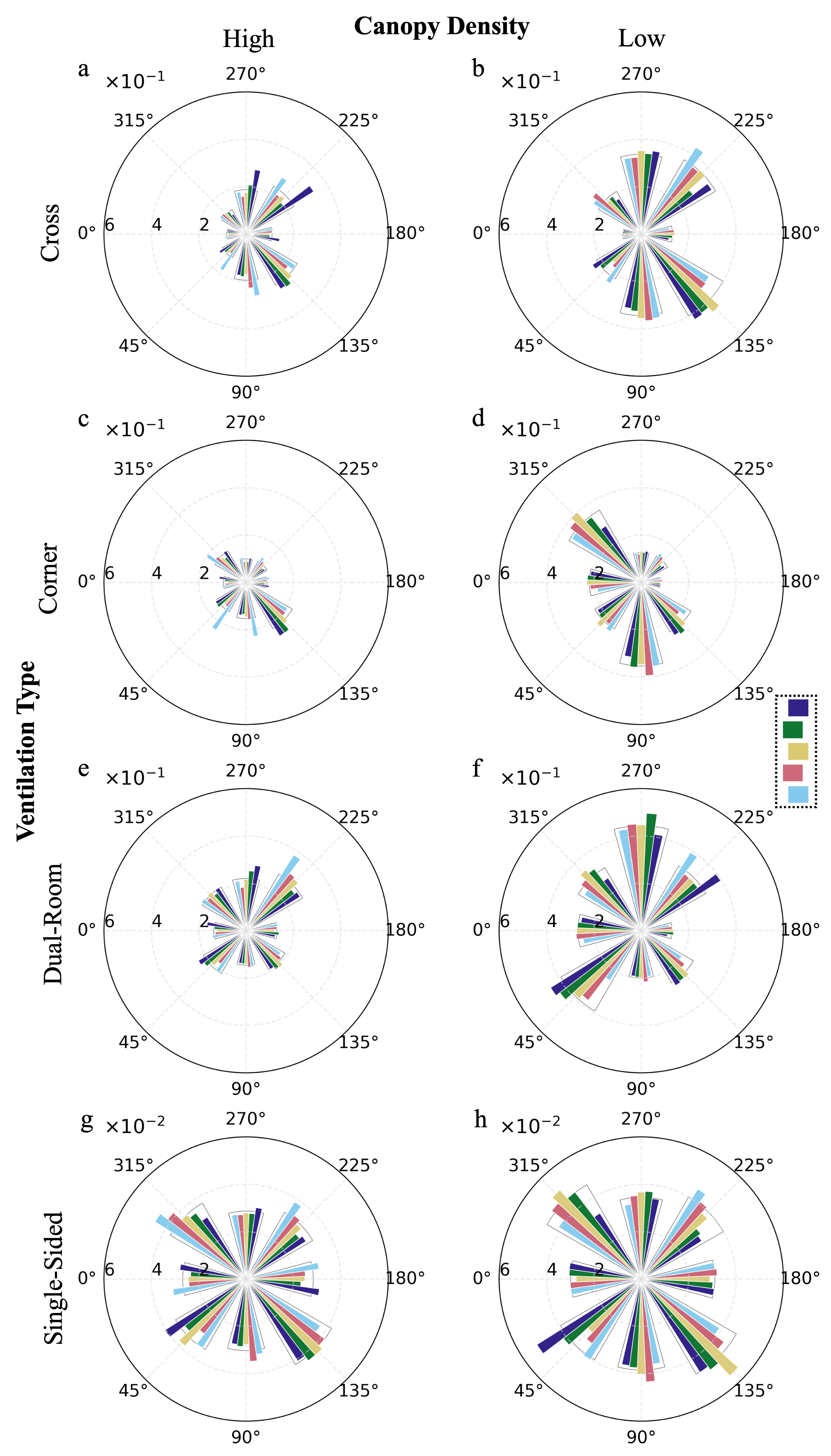}
    \caption{Ventilation roses with eight primary petals, one for each incident wind direction. These primary petals have five sub-petals, one for each house location, with the average ventilation rate silhouetted behind. The sub-petal colors refer to the relative house locations sketched in the color key and highlighted in Figure~\ref{fig:House Locations}. Note the radial axis is 10X smaller for the single sided ventilation.}
    \label{fig:Wind Angles}
\end{figure}
Figure~\ref{fig:Wind Angles} shows eight ventilation roses, one for each room configuration and canopy density. These ventilation roses contain eight primary petals, each with five sub-petals. Like wind roses, the primary petals align with the incident wind angle, with their magnitude (outlined behind the sub-petals) indicating the median ventilation rate. The five colored sub-petals further separate the ventilation by house location, as highlighted by the color key and Figure~\ref{fig:House Locations}. The discussion in this section focuses on the primary petals, whereas the sub-petals will be discussed in Section~\ref{sssec:House Locations}. 

Figure~\ref{fig:Wind Angles} confirms the previous observation regarding the effect of canopy density; the ventilation rate variability due to wind angle is lower for the high-density canopy than for the low-density canopy. However, the wind angles responsible for above-average ventilation rates are mostly the same between the two canopy densities. These wind angles mostly align with the ventilation axis.  

Considering the cross ventilated room first, the wind roses are symmetric, with ventilation rates mirrored across the axis perpendicular to the ventilation axis (\ang{0} and \ang{180}). Higher-than-average ventilation rates are observed for the wind directions that align with the ventilation axis (\ang{90} and \ang{270}). Similarly, high ventilation rates are observed for the \ang{135} and \ang{225} wind directions, where the windows are positioned near the house's upstream corner, and the wind speed outside is higher than that for the \ang{90} and \ang{270} wind directions. In both cases, the flow pattern creates a region of positive stagnation pressure near the inlet window, resulting in higher flow rates. The ventilation rates are significantly lower for wind directions that are perpendicular to the ventilation axis or that place the windows near downstream corners where the pressure assumes a smaller value.

For the corner-ventilated room, the ventilation axis is not aligned with the streets or side yards. As a result, the variability across wind angles becomes more complex, and more pronounced differences between the high and low density canopies appear. In the low-density canopy, the angles that coincide with the ventilation axis (\ang{135} and \ang{315}) both produce higher-than-average ventilation rates. However, there is a significant difference in the flow rates between the two cases, with the \ang{315} direction providing a 50\% higher ventilation rate than the \ang{135} wind direction. This difference, first noted when discussing the flow patterns in Figure~\ref{fig:Local Flow Fields}, can mostly be attributed to differing interference effects for these two wind angles. In the high-density canopy, the interference effects have the opposite effect, and the ventilation flow rate for the \ang{315} wind direction is 50\% higher than for the \ang{135} wind direction. Another notable difference between the high and low density canopy is that, in the low density canopy, the \ang{90} wind angle produces the highest ventilation rate. In this case, the staggered configuration and wider streets support a higher upstream wind velocity, producing a stronger positive stagnation pressure near the upstream window (see Figure~\ref{fig:Flow Fields}).  

Finally, the dual-room setup produces a ventilation rose that is mostly a mirrored version of the corner ventilation rose. This similarity arises because the dual-room geometrically mirrors the corner-ventilated room, with only an additional door opening to the adjacent room. The additional window in the dual-room set-up provides slightly higher ventilation rates across most wind angles, notably for angles where the second room is at the trailing edge of the house with negative relative outdoor pressures. 

In summary, the results indicate that alignment of the wind angle with the ventilation axes provides a useful if incomplete indication of effective ventilation directions. Ultimately, ventilation rates depend on the local flow field. This flow field reflects a combination of wind angle and the positions and geometries of surrounding buildings.

\subsubsection{Sensitivity of Ventilation Rates to House Location} \label{sssec:House Locations}

The five colored sub-petals in Figure~\ref{fig:Wind Angles} visualize the impact of house location within a quadrant. Across house locations, ventilation rates can vary by up to 60\% of the median in the high-density canopy and 50\% in the low-density canopy. %
For parallel flow in the high density canopy, faster flow through the aligned quadrants persists into the first rows of the staggered quadrant. Considering the cross ventilated room, this faster flow results in higher ventilation rates for the most upstream buildings, i.e. the dark blue petal for \ang{270} and the light blue petal for \ang{90}. The corner and dual-room configurations show a similar effect when one of the windows is on the upwind side of the building (\ang{90} for the corner and \ang{270} for the dual-room). In the low-density canopy, this effect is less pronounced. 
With diagonal wind, the large-scale effect of drag anisotropy and the resulting low- and high-velocity streaks create ventilation differences within each quadrant in both the high and low density canopies. This effect is particularly noticeable for the \ang{45} and \ang{225} winds, where the houses near the edge of the quadrant are located in streaks with higher wind speeds and hence have higher ventilation rates for most ventilation configurations.

The most pronounced deviations in ventilation rate occur for homes near the quadrant edges. While these deviations can be partially explained by the large-scale variations in the flow highlighted above, there are cases (e.g., the dual-room set-up at \ang{225}) in which local differences in the canopy geometry upstream or downstream of the edge buildings play a dominant role. The variations in geometry produce local differences in wind direction and wind speed, and we expect that they are common in real urban geometries that have a higher degree of non-uniformity.

\subsection{Implications for natural ventilation modeling} \label{ssec:vent model implecations}

Models for natural ventilation typically determine wind-driven airflow rates based on the pressure differences that drive the flow. These models typically rely on empirical relationships or user-defined inputs for pressure coefficients at window openings~\citep{doe_energy_2025}. These empirical relationships were derived from wind tunnel experiments on isolated rectangular-shaped buildings. If applied, the corrections for interference effects are in the form of a single correction coefficient. For example, Energy Plus's only ventilation model that accounts for a surrounding canopy requires a user-specified shelter class, as defined in the ASHRAE fundamentals handbook. Based on this shelter class, a single factor reduces all wind-driven flow rates~\citep{ashrae_2017_2017, swami_correlations_1988, doe_energy_2025}. %

The results presented in Figure~\ref{fig:Rooms} confirm that, on aggregate, a higher density canopy results in lower ventilation rates. This aggregate observation might seem to support the current approach of applying a factor that reduces ventilation rates for higher density canopies, but the more detailed results presented in Figure~\ref{fig:Wind Angles} indicate that this method does not accurately capture the interference effects that arise in real urban canopies. Changes in flow rates between different canopy densities depend on both the wind angle and surrounding building geometries. Furthermore, different street orientations and related changes in drag introduce large-scale variability in wind speed and direction within a canopy of a given density. The dependency of these local and large-scale effects on specific urban geometry poses a significant modeling challenge given the diverse geometrical configurations in realistic urban areas. To improve accuracy, natural ventilation models must represent both the local and larger-scale effects of a specific urban canopy geometry. Neither of these effects relates well to a single scaling factor applied to the pressure coefficients for an isolated building. Performing large-scale LESs for every building designed to use natural ventilation will be out of reach for the foreseeable future, but there is an opportunity to explore the creation of large representative databases that can inform improved modeling strategies and support the training of data-driven models. %

\section{Conclusion} \label{sec:Conclusion}

This study explored the impact of the urban canopy on wind-driven natural ventilation across four ventilation configurations: cross, corner, dual-room, and single-sided. We performed LES of coupled indoor/outdoor airflow for two canopy densities and varying wind directions. By analyzing the results, we identified the dominant factors that impact natural ventilation flow rates in realistic urban environments.

Flow field visualizations revealed different flow patterns that can affect ventilation rates. At domain-scale, in-canopy variability of wind speeds was observed due to varying $\lambda_f$ and drag anisotropy. Higher $\lambda_f$ slows down canopy flow, while anisotropic canopy drag creates high- and low-velocity streaks. At a more local scale, interference effects from specific canopy elements can significantly alter the ventilation flow into buildings. When adjacent buildings are directly upstream of a window, the wind speed available for ventilation is often significantly reduced. In other scenarios, surrounding buildings can induce local changes in wind angle and wind speed. These changes often reduce ventilation rates but can also increase them by creating more favorable wind angles or amplifying local velocities.

The measured ventilation rates reinforced the effects observed in the flow field visualizations. Median ventilation rates differ by 35\% across multi-window ventilation configurations (cross, corner, and dual-sided), but changes in the surrounding flow field result in significantly more variability in the ventilation rates (50-85\%). For the higher canopy density (with higher $\lambda_f$), the overall reduction in wind speed resulted in 15\% reduced ventilation rates compared to the median. In addition, the flow field has to align more strongly with the streets and side yards, reducing sensitivity to wind angle. Wind angles that align with a room's ventilation axis or place one window on the upwind side of the building often drive high ventilation rates. In both cases, the inflow window is likely to be located in a high-pressure stagnation region, whereas the outflow window will be in a lower pressure region. However, as observed in the flow field visualizations, interference effects from surrounding buildings can play an equally important role in determining ventilation rates for specific wind angles. The importance of the specific configuration of surrounding buildings was further highlighted by considering the ventilation rates obtained in houses near quadrant edges. In certain cases, unique interference effects alter the ventilation rate by up to 60\% compared to houses in quadrant centers. To a lesser extent, the domain-scale effects of varying $\lambda_f$ and drag anisotropy also cause different ventilation rates across houses within a quadrant.

In summary, this study reveals a nuanced interaction between urban canopy geometry and wind angle, which combine to determine the effectiveness of different natural ventilation configurations. Existing models used for the design and analysis of naturally ventilated buildings do not accurately represent these interactions, resulting in significant uncertainty when predicting ventilation rates in new or existing buildings. Future work should focus on developing new models that can better represent these geometry-specific canopy effects to accurately assess natural ventilation and cooling at larger scales. The LES data presented in this paper can provide a starting point for developing such models. Further simulations could explore buoyancy-driven ventilation with openings at different heights, such as skylights, as well as non-neutral boundary layer conditions.

\begin{Backmatter}

\paragraph{Acknowledgements}
Computing for this project was performed on the Sherlock cluster. We would like to thank Stanford University and the Stanford Research Computing Center for providing computational resources and support that contributed to these research results.

\paragraph{Funding Statement}
C.G. gratefully acknowledges funding by a gift from Autodesk Research and Stanford's Strategic Energy Research Consortium.

\paragraph{Declaration of Interests}
The authors declare no conflict of interest.

\paragraph{Author Contributions}
N.B. and C.G. created the research plan. N.B. performed all simulations and data analysis. C.G. and H.S. contributed guidance and advice during the research process. N.B. wrote the manuscript, which H.S. and C.G. reviewed and edited.

\paragraph{Data Availability Statement}
Raw data are available at \url{ https://doi.org/10.25740/vb900bf9392} and from the corresponding author
(N.B.).

\paragraph{Ethical Standards}
The research meets all ethical guidelines, including adherence to the legal requirements of the study country.

\end{Backmatter}

%% file: references.bib
@article{philips_large-eddy_2013,
	title = {Large-eddy simulation of passive scalar dispersion in an urban-like canopy},
	volume = {723},
	issn = {14697645},
	url = {https://www.cambridge.org/core/journals/journal-of-fluid-mechanics/article/largeeddy-simulation-of-passive-scalar-dispersion-in-an-urbanlike-canopy/51F2827D647F1E0EC0B930197FEAB3D4},
	doi = {10.1017/jfm.2013.135},
	journal = {Journal of Fluid Mechanics},
	author = {Philips, D. A. and Rossi, R. and Iaccarino, G.},
	year = {2013},
	note = {Publisher: Cambridge University Press},
	pages = {404--428},
}

@article{grylls_steady-state_2020,
	title = {Steady-{State} {Large}-{Eddy} {Simulations} of {Convective} and {Stable} {Urban} {Boundary} {Layers}},
	volume = {175},
	issn = {1573-1472},
	url = {https://doi.org/10.1007/s10546-020-00508-x},
	doi = {10.1007/s10546-020-00508-x},
	language = {en},
	number = {3},
	urldate = {2025-11-05},
	journal = {Boundary-Layer Meteorology},
	author = {Grylls, Tom and Suter, Ivo and van Reeuwijk, Maarten},
	month = jun,
	year = {2020},
	pages = {309--341},
}

@misc{cascade_technologies_charles_2022,
	title = {{CharLES}},
	url = {https://www.cadence.com/en_US/home/tools/system-analysis/computational-fluid-dynamics/fidelity.html},
	urldate = {2024-09-06},
	author = {{Cascade Technologies}},
	year = {2022},
}

@article{xie_impact_2007,
	title = {Impact of building facades and ground heating on wind flow and pollutant transport in street canyons},
	volume = {41},
	issn = {1352-2310},
	url = {https://www.sciencedirect.com/science/article/pii/S1352231007007121},
	doi = {10.1016/j.atmosenv.2007.08.027},
	number = {39},
	urldate = {2025-11-03},
	journal = {Atmospheric Environment},
	author = {Xie, Xiaomin and Liu, Chun-Ho and Leung, Dennis Y. C.},
	month = dec,
	year = {2007},
	pages = {9030--9049},
}

@article{xue_impact_2024,
	title = {Impact of street canyon morphology on heat and fluid flow: {An} experimental water tunnel study using simultaneous {PIV}-{LIF} technique},
	volume = {150},
	issn = {0894-1777},
	shorttitle = {Impact of street canyon morphology on heat and fluid flow},
	url = {https://www.sciencedirect.com/science/article/pii/S0894177723002224},
	doi = {10.1016/j.expthermflusci.2023.111066},
	urldate = {2025-10-24},
	journal = {Experimental Thermal and Fluid Science},
	author = {Xue, Yunpeng and Zhao, Yongling and Mei, Shuo-Jun and Chao, Yuan and Carmeliet, Jan},
	month = jan,
	year = {2024},
	pages = {111066},
}

@article{hunt_fluid_1999,
	title = {The fluid mechanics of natural ventilation—displacement ventilation by buoyancy-driven flows assisted by wind},
	volume = {34},
	issn = {0360-1323},
	url = {https://www.sciencedirect.com/science/article/pii/S0360132398000535},
	doi = {10.1016/S0360-1323(98)00053-5},
	number = {6},
	urldate = {2025-07-31},
	journal = {Building and Environment},
	author = {Hunt, G. R. and Linden, P. P.},
	month = nov,
	year = {1999},
	pages = {707--720},
}

@misc{california_energy_commission_ca_2022,
	title = {{CA} {Title} 24},
	author = {{California Energy Commission}},
	month = aug,
	year = {2022},
}

@article{ma_inter-scale_2025,
	title = {Inter-scale energy transfer and interaction in a turbulent channel flow with randomly distributed wall roughness},
	volume = {1015},
	issn = {0022-1120, 1469-7645},
	url = {https://www.cambridge.org/core/journals/journal-of-fluid-mechanics/article/interscale-energy-transfer-and-interaction-in-a-turbulent-channel-flow-with-randomly-distributed-wall-roughness/4A770E71C077BBD81276535F968E2488},
	doi = {10.1017/jfm.2025.10352},
	language = {en},
	urldate = {2025-07-23},
	journal = {Journal of Fluid Mechanics},
	author = {Ma, Guo-Zhen and Xu, Chun-Xiao and Sung, Hyung Jin and Tian, Hai-Ping and Huang, Wei-Xi},
	month = jul,
	year = {2025},
	pages = {A2},
}

@book{ashrae_2017_2017,
	address = {Atlanta, Georga},
	edition = {SI edition},
	title = {2017 {ASHRAE} handbook: fundamentals},
	isbn = {978-1-939200-58-7 978-1-5231-1351-4},
	shorttitle = {2017 {ASHRAE} handbook},
	language = {eng},
	publisher = {ASHRAE},
	editor = {{ASHRAE}},
	year = {2017},
}

@misc{doe_energy_2025,
	title = {Energy {Plus} 25.1.0 {Engineering} {Reference}},
	language = {en},
	author = {DOE},
	month = may,
	year = {2025},
}

@article{anderson_numerical_2015,
	title = {Numerical and experimental study of mechanisms responsible for turbulent secondary flows in boundary layer flows over spanwise heterogeneous roughness},
	volume = {768},
	issn = {0022-1120, 1469-7645},
	url = {https://www.cambridge.org/core/journals/journal-of-fluid-mechanics/article/numerical-and-experimental-study-of-mechanisms-responsible-for-turbulent-secondary-flows-in-boundary-layer-flows-over-spanwise-heterogeneous-roughness/859598B7B90B0B28A3E6C32962E8A44D},
	doi = {10.1017/jfm.2015.91},
	language = {en},
	urldate = {2025-07-11},
	journal = {Journal of Fluid Mechanics},
	author = {Anderson, William and Barros, Julio M. and Christensen, Kenneth T. and Awasthi, Ankit},
	month = apr,
	year = {2015},
	pages = {316--347},
}

@article{oke_street_1988,
	title = {Street design and urban canopy layer climate},
	volume = {11},
	issn = {0378-7788},
	url = {https://www.sciencedirect.com/science/article/pii/0378778888900266},
	doi = {10.1016/0378-7788(88)90026-6},
	number = {1},
	urldate = {2025-06-14},
	journal = {Energy and Buildings},
	author = {Oke, T. R.},
	month = mar,
	year = {1988},
	pages = {103--113},
}

@article{hussain_wind_1980,
	title = {A wind tunnel study of the mean pressure forces acting on large groups of low-rise buildings},
	volume = {6},
	issn = {0167-6105},
	url = {https://www.sciencedirect.com/science/article/pii/0167610580900021},
	doi = {10.1016/0167-6105(80)90002-1},
	number = {3},
	urldate = {2025-06-13},
	journal = {Journal of Wind Engineering and Industrial Aerodynamics},
	author = {Hussain, M. and Lee, B. E.},
	month = oct,
	year = {1980},
	pages = {207--225},
}

@incollection{ambo_aerodynamic_2020,
	series = {{AIAA} {SciTech} {Forum}},
	title = {Aerodynamic force prediction of the laminar to turbulent flow transition around the front bumper of the vehicle using {Dynamic}-slip wall model {LES}},
	url = {https://arc.aiaa.org/doi/10.2514/6.2020-0036},
	urldate = {2025-01-18},
	booktitle = {{AIAA} {Scitech} 2020 {Forum}},
	publisher = {American Institute of Aeronautics and Astronautics},
	author = {Ambo, Kei and Nagaoka, Hiroaki and Philips, David A. and Ivey, Chris and Brès, Guillaume A. and Bose, Sanjeeb T.},
	month = jan,
	year = {2020},
	doi = {10.2514/6.2020-0036},
}

@article{chen_full-scale_2022,
	title = {Full-scale validation of {CFD} simulations of buoyancy-driven ventilation in a three-story office building},
	volume = {221},
	issn = {03601323},
	url = {https://linkinghub.elsevier.com/retrieve/pii/S0360132322004759},
	doi = {10.1016/j.buildenv.2022.109240},
	language = {en},
	urldate = {2024-10-31},
	journal = {Building and Environment},
	author = {Chen, Chen and Gorlé, Catherine},
	month = aug,
	year = {2022},
	pages = {109240},
}

@article{aboshosha_abl_2015,
	title = {{LES} of {ABL} flow in the built-environment using roughness modeled by fractal surfaces},
	volume = {19},
	issn = {2210-6707},
	url = {https://www.sciencedirect.com/science/article/pii/S2210670715300056},
	doi = {10.1016/j.scs.2015.07.003},
	urldate = {2024-10-31},
	journal = {Sustainable Cities and Society},
	author = {Aboshosha, Haitham and Bitsuamlak, Girma and Damatty, Ashraf El},
	month = dec,
	year = {2015},
	pages = {46--60},
}

@article{vreman_eddy-viscosity_2004,
	title = {An eddy-viscosity subgrid-scale model for turbulent shear flow: {Algebraic} theory and applications},
	volume = {16},
	issn = {1070-6631, 1089-7666},
	shorttitle = {An eddy-viscosity subgrid-scale model for turbulent shear flow},
	url = {https://pubs.aip.org/pof/article/16/10/3670/255289/An-eddy-viscosity-subgrid-scale-model-for},
	doi = {10.1063/1.1785131},
	language = {en},
	number = {10},
	urldate = {2024-10-25},
	journal = {Physics of Fluids},
	author = {Vreman, A. W.},
	month = oct,
	year = {2004},
	pages = {3670--3681},
}

@article{hochschild_comparison_2024,
	title = {Comparison of measured and {LES}-predicted wind pressures on the {Space} {Needle}},
	volume = {249},
	issn = {0167-6105},
	url = {https://www.sciencedirect.com/science/article/pii/S0167610524001120},
	doi = {10.1016/j.jweia.2024.105749},
	urldate = {2024-10-25},
	journal = {Journal of Wind Engineering and Industrial Aerodynamics},
	author = {Hochschild, John and Gorlé, Catherine},
	month = jun,
	year = {2024},
	pages = {105749},
}

@article{tong_mapping_2021,
	title = {Mapping the urban natural ventilation potential by hydrological simulation},
	url = {https://doi.org/10.1007/s12273-020-0755-6},
	doi = {10.1007/s12273-020-0755-6},
	journal = {Building Simulation},
	author = {Tong, Ziyu and Luo, Yu and Zhou, Juelun},
	year = {2021},
}

@article{ciarlatani_investigation_2023,
	title = {Investigation of peak wind loading on a high-rise building in the atmospheric boundary layer using large-eddy simulations},
	volume = {236},
	issn = {0167-6105},
	url = {https://www.sciencedirect.com/science/article/pii/S0167610523001113},
	doi = {10.1016/j.jweia.2023.105408},
	urldate = {2024-07-10},
	journal = {Journal of Wind Engineering and Industrial Aerodynamics},
	author = {Ciarlatani, Mattia Fabrizio and Huang, Zhu and Philips, David and Gorlé, Catherine},
	month = may,
	year = {2023},
	pages = {105408},
}

@article{hwang_large-eddy_2022,
	title = {Large-{Eddy} {Simulations} of {Wind}-{Driven} {Cross} {Ventilation}, {Part} 2: {Comparison} of {Ventilation} {Performance} {Under} {Different} {Ventilation} {Configurations}},
	volume = {8},
	issn = {2297-3362},
	shorttitle = {Large-{Eddy} {Simulations} of {Wind}-{Driven} {Cross} {Ventilation}, {Part} 2},
	url = {https://www.frontiersin.org/articles/10.3389/fbuil.2022.911253},
	urldate = {2023-12-07},
	journal = {Frontiers in Built Environment},
	author = {Hwang, Yunjae and Gorlé, Catherine},
	year = {2022},
}

@article{hwang_large-eddy_2022-1,
	title = {Large-{Eddy} {Simulations} of {Wind}-{Driven} {Cross} {Ventilation}, {Part1}: {Validation} and {Sensitivity} {Study}},
	volume = {8},
	issn = {2297-3362},
	shorttitle = {Large-{Eddy} {Simulations} of {Wind}-{Driven} {Cross} {Ventilation}, {Part1}},
	url = {https://www.frontiersin.org/articles/10.3389/fbuil.2022.911005},
	urldate = {2023-11-16},
	journal = {Frontiers in Built Environment},
	author = {Hwang, Yunjae and Gorlé, Catherine},
	year = {2022},
}

@techreport{orme_applicable_1999,
	title = {Applicable models for air infiltration and ventilation calculations},
	language = {en},
	institution = {International Energy Agency},
	author = {Orme, Malcolm},
	year = {1999},
}

@book{holmes_wind_2018,
	address = {Boca Raton, FL},
	edition = {Third edition},
	title = {Wind loading of structures},
	isbn = {978-1-4822-2922-6},
	language = {eng},
	publisher = {CRC Press},
	author = {Holmes, John D.},
	year = {2018},
	note = {OCLC: 1275089035},
}

@book{stull_introduction_2003,
	address = {Dordrecht},
	edition = {Reprint},
	series = {Atmospheric sciences library},
	title = {An introduction to boundary layer meteorology},
	isbn = {978-94-009-3027-8 978-90-277-2769-5},
	language = {eng},
	number = {13},
	publisher = {Kluwer},
	author = {Stull, Roland B.},
	year = {2003},
}

@article{adachi_numerical_2020,
	title = {Numerical simulation for cross-ventilation flow of generic block sheltered by urban-like block array},
	volume = {185},
	issn = {0360-1323},
	url = {https://www.sciencedirect.com/science/article/pii/S0360132320305473},
	doi = {10.1016/j.buildenv.2020.107174},
	urldate = {2023-08-31},
	journal = {Building and Environment},
	author = {Adachi, Y. and Ikegaya, N. and Satonaka, H. and Hagishima, A.},
	month = nov,
	year = {2020},
	pages = {107174},
}

@article{swami_correlations_1988,
	title = {Correlations for pressure distribution on buildings and calculation of natural-ventilation airflow},
	url = {https://www.semanticscholar.org/paper/Correlations-for-pressure-distribution-on-buildings-Swami-Chandra/8f606c922d4792e1a5ba748c95f16a896ecf4e8e},
	urldate = {2023-08-30},
	journal = {Ashrae Transactions},
	author = {Swami, M. and Chandra, S.},
	year = {1988},
}

@article{king_investigating_2017,
	title = {Investigating the influence of neighbouring structures on natural ventilation potential of a full-scale cubical building using time-dependent {CFD}},
	volume = {169},
	issn = {0167-6105},
	url = {https://www.sciencedirect.com/science/article/pii/S016761051730257X},
	doi = {10.1016/j.jweia.2017.07.020},
	urldate = {2023-08-31},
	journal = {Journal of Wind Engineering and Industrial Aerodynamics},
	author = {King, Marco-Felipe and Gough, Hannah L. and Halios, Christos and Barlow, Janet F. and Robertson, Adam and Hoxey, Roger and Noakes, Catherine J.},
	month = oct,
	year = {2017},
	pages = {265--279},
}

@article{van_hooff_accuracy_2017,
	title = {On the accuracy of {CFD} simulations of cross-ventilation flows for a generic isolated building: {Comparison} of {RANS}, {LES} and experiments},
	volume = {114},
	issn = {0360-1323},
	shorttitle = {On the accuracy of {CFD} simulations of cross-ventilation flows for a generic isolated building},
	url = {https://www.sciencedirect.com/science/article/pii/S036013231630511X},
	doi = {10.1016/j.buildenv.2016.12.019},
	urldate = {2023-08-30},
	journal = {Building and Environment},
	author = {van Hooff, T. and Blocken, B. and Tominaga, Y.},
	month = mar,
	year = {2017},
	pages = {148--165},
}

@article{ramponi_energy_2014,
	title = {Energy saving potential of night ventilation: {Sensitivity} to pressure coefficients for different {European} climates},
	volume = {123},
	issn = {03062619},
	shorttitle = {Energy saving potential of night ventilation},
	url = {https://linkinghub.elsevier.com/retrieve/pii/S0306261914001858},
	doi = {10.1016/j.apenergy.2014.02.041},
	language = {en},
	urldate = {2023-08-30},
	journal = {Applied Energy},
	author = {Ramponi, Rubina and Angelotti, Adriana and Blocken, Bert},
	month = jun,
	year = {2014},
	pages = {185--195},
}

@article{costola_overview_2009,
	title = {Overview of pressure coefficient data in building energy simulation and airflow network programs},
	volume = {44},
	issn = {03601323},
	url = {https://linkinghub.elsevier.com/retrieve/pii/S0360132309000444},
	doi = {10.1016/j.buildenv.2009.02.006},
	language = {en},
	number = {10},
	urldate = {2023-08-30},
	journal = {Building and Environment},
	author = {Cóstola, D. and Blocken, B. and Hensen, J.L.M.},
	month = oct,
	year = {2009},
	pages = {2027--2036},
}

@article{hwang_large-eddy_2023,
	title = {Large-eddy simulations to define building-specific similarity relationships for natural ventilation flow rates},
	volume = {3},
	doi = {10.1017/flo.2023.4},
	journal = {Flow},
	author = {Hwang, Yunjae and Gorlé, Catherine},
	month = mar,
	year = {2023},
}

@article{lu_novel_2023,
	title = {Novel {Geometric} {Parameters} for {Assessing} {Flow} {Over} {Realistic} {Versus} {Idealized} {Urban} {Arrays}},
	volume = {15},
	copyright = {© 2023 The Authors. Journal of Advances in Modeling Earth Systems published by Wiley Periodicals LLC on behalf of American Geophysical Union.},
	issn = {1942-2466},
	url = {https://onlinelibrary.wiley.com/doi/abs/10.1029/2022MS003287},
	doi = {10.1029/2022MS003287},
	language = {en},
	number = {7},
	urldate = {2023-08-16},
	journal = {Journal of Advances in Modeling Earth Systems},
	author = {Lu, Jiachen and Nazarian, Negin and Hart, Melissa Anne and Krayenhoff, E. Scott and Martilli, Alberto},
	year = {2023},
	note = {\_eprint: https://onlinelibrary.wiley.com/doi/pdf/10.1029/2022MS003287},
	pages = {e2022MS003287},
}

@article{freire_improvement_2013,
	title = {On the improvement of natural ventilation models},
	volume = {62},
	issn = {0378-7788},
	url = {https://www.sciencedirect.com/science/article/pii/S0378778813001497},
	doi = {10.1016/j.enbuild.2013.02.055},
	language = {en},
	urldate = {2023-07-29},
	journal = {Energy and Buildings},
	author = {Freire, Roberto Z. and Abadie, Marc O. and Mendes, Nathan},
	month = jul,
	year = {2013},
	pages = {222--229},
}

@article{chiesa_python-based_2019,
	title = {Python-based calculation tool of wind-pressure coefficients on building envelopes},
	volume = {1343},
	issn = {1742-6588, 1742-6596},
	url = {https://iopscience.iop.org/article/10.1088/1742-6596/1343/1/012132},
	doi = {10.1088/1742-6596/1343/1/012132},
	language = {en},
	number = {1},
	urldate = {2023-07-29},
	journal = {Journal of Physics: Conference Series},
	author = {Chiesa, Giacomo and Grosso, Mario},
	month = nov,
	year = {2019},
	pages = {012132},
}

@article{xue_quantifying_2022,
	title = {Quantifying the spatial homogeneity of urban road networks via graph neural networks},
	volume = {4},
	copyright = {2022 The Author(s), under exclusive licence to Springer Nature Limited},
	issn = {2522-5839},
	url = {https://www.nature.com/articles/s42256-022-00462-y},
	doi = {10.1038/s42256-022-00462-y},
	language = {en},
	number = {3},
	urldate = {2023-07-11},
	journal = {Nature Machine Intelligence},
	author = {Xue, Jiawei and Jiang, Nan and Liang, Senwei and Pang, Qiyuan and Yabe, Takahiro and Ukkusuri, Satish V. and Ma, Jianzhu},
	month = mar,
	year = {2022},
	note = {Number: 3
Publisher: Nature Publishing Group},
	pages = {246--257},
}

@article{biljecki_global_2022,
	title = {Global {Building} {Morphology} {Indicators}},
	volume = {95},
	issn = {0198-9715},
	url = {https://www.sciencedirect.com/science/article/pii/S0198971522000539},
	doi = {10.1016/j.compenvurbsys.2022.101809},
	language = {en},
	urldate = {2023-07-02},
	journal = {Computers, Environment and Urban Systems},
	author = {Biljecki, Filip and Chow, Yoong Shin},
	month = jul,
	year = {2022},
	pages = {101809},
}

@article{adolphe_simplified_2001,
	title = {A {Simplified} {Model} of {Urban} {Morphology}: {Application} to an {Analysis} of the {Environmental} {Performance} of {Cities}},
	volume = {28},
	issn = {0265-8135, 1472-3417},
	shorttitle = {A {Simplified} {Model} of {Urban} {Morphology}},
	url = {http://journals.sagepub.com/doi/10.1068/b2631},
	doi = {10.1068/b2631},
	language = {en},
	number = {2},
	urldate = {2023-07-02},
	journal = {Environment and Planning B: Planning and Design},
	author = {Adolphe, Luc},
	month = apr,
	year = {2001},
	pages = {183--200},
}

@article{letzel_high_2008,
	title = {High resolution urban large-eddy simulation studies from street canyon to neighbourhood scale},
	volume = {42},
	issn = {1352-2310},
	url = {https://www.sciencedirect.com/science/article/pii/S1352231008007036},
	doi = {10.1016/j.atmosenv.2008.08.001},
	language = {en},
	number = {38},
	urldate = {2023-06-21},
	journal = {Atmospheric Environment},
	author = {Letzel, Marcus Oliver and Krane, Martina and Raasch, Siegfried},
	month = dec,
	year = {2008},
	pages = {8770--8784},
}

@article{hirose_indoor_2021,
	title = {Indoor airflow and thermal comfort in a cross-ventilated building within an urban-like block array using large-eddy simulations},
	volume = {196},
	url = {https://doi.org/10.1016/j.buildenv.2021.107811},
	doi = {10.1016/j.buildenv.2021.107811},
	journal = {Building and Environment},
	author = {Hirose, C and Ikegaya, N and Hagishima, A and Tanimoto, J},
	year = {2021},
	pages = {107811},
}

@article{van_nguyen_new_2018,
	title = {New surrogate model for wind pressure coefficients in a schematic urban environment with a regular pattern},
	volume = {9},
	issn = {20734433},
	doi = {10.3390/ATMOS9030113},
	number = {3},
	urldate = {2022-03-21},
	journal = {Atmosphere},
	author = {Van Nguyen, Tam and De Troyer, Frank},
	month = mar,
	year = {2018},
	note = {Publisher: MDPI AG},
}

@article{zawadzka_assessment_2021,
	title = {Assessment of heat mitigation capacity of urban greenspaces with the use of {InVEST} urban cooling model, verified with day-time land surface temperature data},
	volume = {214},
	issn = {01692046},
	doi = {10.1016/j.landurbplan.2021.104163},
	journal = {Landscape and Urban Planning},
	author = {Zawadzka, J. E. and Harris, J. A. and Corstanje, R.},
	month = oct,
	year = {2021},
	note = {Publisher: Elsevier B.V.},
}

@techreport{dean_future_2018,
	title = {The {Future} of {Cooling}},
	url = {www.iea.org/t&c/},
	institution = {International Energy Agency},
	author = {Dean, Brian and Dulac, John and Morgan, Trevor and Remme, Uwe},
	year = {2018},
}
